\documentclass[preprint,authoryear,5p]{elsarticle}

\usepackage{amssymb}

\biboptions{comma,round}

\pdfoutput=1

\usepackage{graphicx}
\usepackage[space]{grffile}
\usepackage{latexsym}
\usepackage{amsfonts,amsmath,amssymb}
\usepackage{url}
\usepackage[utf8]{inputenc}
\usepackage{fancyref}
\usepackage{hyperref}
\hypersetup{colorlinks=false,pdfborder={0 0 0},}
\usepackage{courier}
\usepackage{float}
\usepackage{booktabs}
\usepackage{threeparttable}
\usepackage{array}
\usepackage{tabu}

\usepackage{comment}

\usepackage[usenames,dvipsnames,svgnames,table]{xcolor}

\newcommand{\aj}{AJ}			
\newcommand{\araa}{ARA\&A}		
\newcommand{\apj}{ApJ}			
\newcommand{\apjl}{ApJLett}		
\newcommand{\apjs}{ApJS}		
\newcommand{\apss}{Ap\&SS}		
\newcommand{\aap}{A\&A}			
\newcommand{\mnras}{MNRAS}		

\usepackage{relsize}
\newcommand\Cpp{C\nolinebreak[4]\hspace{-.05em}\raisebox{.4ex}{\relsize{-3}{\textbf{++}}}}

\definecolor{midgray}{gray}{0.95}%

\usepackage{listings}

\lstdefinestyle{code}{%
     basicstyle=\footnotesize\ttfamily,       
     frame=single,               
     framesep=2pt,               
     framerule=0.8pt,            
     rulecolor=\color{midgray},  
     breaklines=true,           
     breakindent=0pt,
     backgroundcolor=\color{midgray},
     prebreak=\mbox{\textbackslash{}},
     captionpos=b, 
     abovecaptionskip=15pt          
}

\lstdefinestyle{python}{
  style=code,
  language=Python,
  showstringspaces=false,
  keywordstyle=\bfseries\color{green!40!black},
  commentstyle=\itshape\color{gray},
  identifierstyle=\color{black},
  stringstyle=\color{gray},
  rulecolor=\color{midgray}, 
}

\lstdefinestyle{java}{
  style=python,
  language=Java
}

\journal{Astronomy and Computing}

\begin{document}

\begin{frontmatter}



\title{Iris: an Extensible Application for Building and Analyzing\\
Spectral Energy Distributions}

\author[sao]{Omar Laurino~\corref{cor}}
\cortext[cor]{Corresponding author}
\ead{olaurino@cfa.harvard.edu}

\author[sao]{Jamie Budynkiewicz}

\author[sao]{Raffaele D'Abrusco}

\author[sao]{Nina Bonaventura~\fnref{nina}}
\fntext[nina]{Present affiliation: McGill University, 3600 University St. Montr\'{e}al QC, Canada H3A 2T8}

\author[stsci]{Ivo Busko}

\author[sao]{Mark Cresitello-Dittmar}

\author[sao]{Stephen M. Doe~\fnref{doe}}
\fntext[doe]{Present address: 4 Lafayette Drive, East Walpole, MA 02032}

\author[ipac]{Rick Ebert}

\author[sao]{Janet D. Evans}

\author[noao]{Patrick Norris}

\author[ipac]{Olga Pevunova}

\author[sao]{Brian Refsdal}

\author[noao]{Brian Thomas}

\author[stsci]{Randy Thompson}

\address[sao]{Smithsonian Astrophysical Observatory, 60 Garden St.
Cambridge, MA 02138}
\address[stsci]{Space Telescope Science Institute, 3700 San Martin Dr.
Baltimore, MD 21218}
\address[ipac]{Infrared Processing and Analysis Center, 770 South Wilson Ave.
Pasadena, CA 91125}
\address[noao]{National Optical Astronomy Observatory, 950 N Cherry Ave.
Tucson, AZ 85719}

\begin{abstract}
Iris is an extensible application that provides astronomers with a user-friendly
interface capable of ingesting broad-band data from many different sources
in order to build, explore, and model spectral energy distributions (SEDs). Iris
takes advantage of the standards defined by the International Virtual
Observatory Alliance, but hides the technicalities of such standards by
implementing different layers of abstraction on top of them. Such intermediate
layers provide hooks that users and developers can exploit in order to extend
the capabilities provided by Iris. For instance, custom Python models can be
combined in arbitrary ways with the Iris built-in models or with other custom
functions. As such, Iris offers a platform for the development and integration
of SED data, services, and applications, either from the user's system or from
the web. In this paper we describe the built-in features provided by Iris for
building and analyzing SEDs. We also explore in some detail the Iris framework
and software development kit, showing how astronomers and software developers
can plug their code into an integrated SED analysis environment.  \end{abstract}


\end{frontmatter}

\bibliographystyle{model2-names}

\label{sec:introduction} \section{Introduction} 

The emission processes of astronomical objects (e.g., stars, galaxies, quasars)
are reflected in the spectral energy distribution (SED) of the radiation detected
by astronomers with a variety of telescopes and instruments. Astronomers use this
information to infer the physical properties of the source by comparing the
detected SED with different emission models. Methods for these studies have
been developed by several communities in astronomy, focused on either a particular
type of source, or quite often on a particular region of the emission spectrum
(e.g., radio, optical-IR, X-ray). These focused tools typically require different
input formats and imply the use of wavelength specific units, as well as being
optimized for particular models to compare the SED with. However, the most
complete picture of any emission phenomena requires the use of the most complete
information base. Modern wide-field ground and space telescopes, and the
availability of data from multi-wavelenth archives, allow in principle to build
and study broadband SEDs for any kind of astronomical object. However, a tool that can
efficiently and powerfully make use of this information requires a 
non-wavelength-specific approach.

\begin{sloppypar}
The International Virtual
Observatory Alliance \citep[IVOA;][]{2004SPIE.5493..137Q} provides a set of
standards and protocols that facilitate interoperability among astronomy-related
services and tools. These IVOA specifications can be implemented to enable
generalized SED analysis, regardless of the spectral regime
and objects being studied.
\end{sloppypar}

\begin{sloppypar}
In order to design effective applications, one wants
to leverage IVOA standards without exposing the complexity and
technicality of their specifications to the users. Also, while application
developers implement many desired features, it is useful, and sometimes 
required, to provide hooks for users and third party developers to extend
the application's functionality without requiring knowledge of standards themselves.
Designing such an application, like a general SED analysis tool, thus becomes an
exercise in designing a framework that implements some basic,
effective functionality for a wide set of use cases, while being highly extensible.
\bigskip

\textit{Iris}, the Virtual Astronomical Observatory \citep[VAO;][]{2012SPIE.8449E..0HB}
SED analysis tool, is such an IVOA-enabled desktop application. 
With Iris, users may populate SEDs with data from files, built-in portals to data
archives, and other Virtual Observatory (VO) applications. Users can
interactively visualize and edit SEDs, and fit SEDs with fine-tuned modeling
features. Iris provides a suite of astrophysical models, but also lets users
import custom models and template libraries. All front-end features
of Iris completely hide the underlying technical IVOA specifications from
the user.

While implementing IVOA standards and protocols, 
we took advantage of existing astronomy
software, namely
Specview~\citep{2002ASPC..281..120B} for the visualization and fitting user
interfaces, the NASA/IPAC Extragalactic Database (NED) SED 
Service\footnote{\url{http://vo.ned.ipac.caltech.edu/SED_Service/}} for data
acquisition, and Sherpa~\citep{2001SPIE.4477...76F,2009pysc.conf...51R} for the 
modeling and fitting engine. Along with these components, new ones, like the SED 
Builder, were developed specifically for Iris
\citep{2012ASPC..461..893D,2013AAS...22124038L}.
\end{sloppypar}




Iris was developed inside the framework of the VAO science applications: the
different components were contributed by developers from the Smithsonian
Astrophysical Observatory, the Space Telescope Science Institute (STScI), and the NASA
Infrared Processing and Analysis Center (IPAC). Quality assurance and testing were led
by team members at the National Optical Astronomy Observatory and STScI.


In this paper we present the Iris application, design, and extensible architecture. In Section
\ref{sec:overview} we briefly explore the landscape of SED applications and
analysis tools that Iris joined, and provide an example use-case of Iris. We
explore how astronomers can include their own models or templates as Python
functions in Section \ref{sec:usermodels}. An introduction to Iris' general
architecture (the Iris \emph{stack}) is illustrated in Section \ref{sec:stack}.
A more detailed overview of the Iris extensible framework design (Section
\ref{sec:architecture}) is followed by a detailed description of the more
advanced Iris capabilities (Section \ref{sec:components}). Finally, we
describe the Iris software development kit, including a ``How-to'' on extending
Iris with plug-ins (Section \ref{sec:plugins}). Sections \ref{sec:architecture} and 
\ref{sec:plugins} are targeted to software
developers.

The paper refers to version 2.0.1 of Iris. Iris can be downloaded as a binary
archive for OS X and
Linux\footnote{\url{http://cxc.cfa.harvard.edu/iris/latest/download/}}, and the
source code is hosted on GitHub as a public
repository\footnote{\url{https://github.com/ChandraCXC/iris}}.

\section{SED Analysis with Iris} \label{sec:overview}

Fitting spectral energy distributions enables astron\-omers to estimate fundamental 
properties of various astronomical objects. 
In galaxy evolution studies, for example, stellar mass, star formation rates, dust content, and redshift
are often derived from galaxy SEDs (e.g.\citet{1998AJ....115.1329S},
\citet{2001ApJS..137..139S}, \citet{2007ApJS..169..328R}, and many others). 
Accretion disks surrounding supermassive black holes, x-ray binary and young stellar objects can be studied by
fitting models to the host objects' SEDs, extracting information like 
accretion rates, disk geometry, and disk temperature \citep[e.g.,][]{1987ApJ...321..305C,1990A&A...235..162V,1997ApJ...490..368C,2006ApJS..167..256R}. 
Stellar SED analysis can recognize mid IR excess, which may indicate circumstellar 
disks~\citep{2000prpl.conf..639L,2005ApJ...623..493C}. As these examples show, SEDs 
are widely used throughout astronomy.



With ever increasing wide-field surveys and datasets over the years,
astronomers have been able to use multi-wavelength SEDs more frequently for
their research. As such, many robust SED analysis codes have been created to
help astronomers mod\-el, fit, and derive physical quantities from SEDs
\citep{2011Ap&SS.331....1W,2013ARA&A..51..393C}. These widely-used codes
implement a diverse set of methods, for instance: inversion (e.g., 
STARLIGHT~[\citealp{2004MNRAS.355..273C}] and
PAHFIT~[\citealp{2007ApJ...656..770S}]),
principal component analysis~\citep[e.g.,][]{2009MNRAS.394.1496B},
$\mathrm{\chi}^{2}$-minimization codes 
(e.g., Le PHARE~[\citealp{1999MNRAS.310..540A,2006A&A...457..841I}] and 
HyperZ~[\citealp{2000A&A...363..476B}]), and Bayesian inference 
(e.g., BPZ~[\citealp{2000ApJ...536..571B}], 
VOSA~[\citealp{2008A&A...492..277B}], and 
GalMC~[\citealp{2011ApJ...737...47A}]).

Most widely used fitting packages are tailored for specific data sets or
spectral ranges (such as PAHFIT and STAR\-LIGHT), providing robust
fitting methods and results. They require the data to be in a specific format
with specific units in order for the tool to work properly. When fitting a
broadband SED that spans over decades in the spectrum, the astronomer will
typically gather datasets from different public archives and colleagues in order
to add such data to their own. More often than not, the datasets are stored in
different file formats and units. 
The user must provide their own methods to extract the necessary data from each
file, homogenize the units, and output a file in the format supported by the
tool; converting the data to a supported format may easily become a tedious task
with each additional dataset.

\begin{table*}[tp!]
\caption{\textbf{Supported file formats.} Native formats are automatically loaded into
Iris. Supported formats require some user input to map the file data to the spectral
and flux information.}
\label{table:file_formats}
\centering
\begin{threeparttable}
\begin{tabular}{llm{14cm}}
\toprule
          & Format      & Description \\
\cmidrule{2-3}
          & VOTable     & XML-based format, text or binary following IVOA Spectrum Data Model v1.0, 1.1, or 1.2. \\
\rotatebox{90}{\rlap{Native~}}
          & FITS        & Series of HDUs\tnote{a} ~with text header and text or binary data extensions following IVOA Spectrum Data Model v1.0, 1.1, or 1.2. \\
\cmidrule{2-3}
          & VOTable     & XML-format, text or binary. \\
          & FITS        & Series of HDUs\tnote{a} ~with text header and text or binary data extensions. \\
          & ASCII Table & Text file with columns separated by spaces and/or tabs. \\
          & CSV         & Text file with columns separated by commas (first row may contain column names). \\
\rotatebox{90}{\rlap{~Supported}}
          & IPAC        & A custom bar-separated text format by IPAC. \\
          & TST         & Tab Separated Table (comments are ignored, metadata is in key, value pairs). \\
\bottomrule                                                                         
\end{tabular}
\begin{tablenotes}
\item [a] \footnotesize Header Data Units.
\end{tablenotes}
\end{threeparttable}

\end{table*}

\begin{table}[h!]
\caption{\textbf{Supported SED units.} Iris can read, write, and/or plot data in
the spectral and flux units listed in this table. Italicized units are only available
for plotting.}
\label{table:units}
\centering
\begin{threeparttable}
\begin{tabular}{ll}
    \toprule
    Spectral Axis                & Flux Axis \\
    \midrule
    \mbox{\AA}              & $\mathrm{erg}/\mathrm{s}/\mathrm{cm}^{2}/\mbox{\AA}$     \\
    nm                      & $\mathrm{erg}/\mathrm{s}/\mathrm{cm}^{2}/\mathrm{Hz}$    \\
    $\mathrm{\mu}$m         & $\mathrm{photon}/\mathrm{s}/\mathrm{cm}^{2}/\mbox{\AA}$  \\
    \textit{mm}             & $\mathrm{photon}/\mathrm{s}/\mathrm{cm}^{2}/\mathrm{Hz}$ \\
    cm                      & $\mathrm{Watt}/\mathrm{m}^{2}/{\mu}\mathrm{m}$           \\
    m                       & \textit{$Watt/cm^{2}/{\mu}m$}            \\
    eV                      & \textit{$Watt/m^{2}/nm$}               \\
    keV                     & $\mathrm{Watt}/\mathrm{m}^{2}/\mathrm{Hz}$               \\
    MeV                     & \textit{$Rayleigh/\mbox{\AA}$}                           \\
    Hz                      & Jy                									   \\
    kHz                     & mJy               									   \\
    MHz                     & \textit{${\mu}$Jy}       								   \\
    GHz                     & ~                 									   \\
    THz                     & AB mag           										   \\
    \textit{1/${\mu}$m}     & ST mag\tnote{a}      									   \\
    \textit{km/s @ 21 cm}   & ~               										   \\
    \textit{km/s @ 12 CO}   & Jy Hz             									   \\
    ~ 					    & $\mathrm{erg}/\mathrm{s}/\mathrm{cm}^{2}$                \\
    \bottomrule
\end{tabular}
\begin{tablenotes}
\item [a] \footnotesize$\mathrm{ST = -2.5~log_{10}}~f_{\lambda} - 21.10$, where $f_{\lambda}$ is
          the source flux density expressed per unit wavelength.
\end{tablenotes}
\end{threeparttable}

\end{table}

While SED analysis tools often have different input formats from each other,
they effectively require the same information to run. Whether datasets are
stored in a FITS file, a tab-separated ASCII table, or a 
VOTable~\citep{2011arXiv1110.0524O} coming from a
VO data discovery application, they are all serializations of the same, global,
abstract, scientific model of photometric measurements for
astronomical sources.

By employing a standardized definition of such models, Iris streamlines
the process of building SEDs for analysis. In other terms, one of the goals of
Iris is to make SED building a painless and straightforward process, letting the
scientist focus on the sophisticated and original parts of the scientific
work-flow: data analysis, hypothesis testing, and knowledge extraction.

Following VO efforts to combine data services and applications seamlessly, Iris
offers an interface for building large broadband SEDs from different sources in
various data formats, while providing robust fitting methods and interactive
visualization capabilities using existing astronomical software.
It is important to stress that this is not only a matter of ingesting
non-standard files, but also to allow scientists to create standardized versions
of their datasets: the improved interoperability enables more tools, inside or
outside Iris, to load and interpret such datasets with minimal user
intervention.

\begin{sloppypar}
Much effort has been put into making Iris lenient on data format. While natively
supporting VO-compliant files (properly annotated
VOTable and FITS files), Iris can ingest ASCII, CSV,
and other table-like formats as well with some extra user input.
Table~\ref{table:file_formats}
describes the file formats that can be read into Iris. Users may also
seamlessly transfer data from other VO applications or data archive services
through SAMP, the Simple Application Messaging
Protocol~\citep{2011arXiv1110.0528T}. Moreover, Iris can read, write, and display SED data in
a variety of commonly-used units, which are listed in Table~\ref{table:units}, with minimal user effort.

\end{sloppypar}

But more importantly, Iris provides standardized views of the integrated
datasets to its clients, whether they are built-in components, third party
plug-ins, or external applications.

\subsection{A Use Case} \label{subsec:usecase}

In this section, we present a brief, illustrative use-case
of Iris to showcase its main features. We outline the analysis of the broadband SED
of flat spectrum radio quasar (FSRQ) object PKS 1127-14~\citep[see][]{2004ApJ...600L..27B}, and save the
results to file.

For details on the Iris features introduced in this use-case, see Section~\ref{sec:components}.

\begin{figure*} \centering
\includegraphics[height=2.95in,width=5.3in]{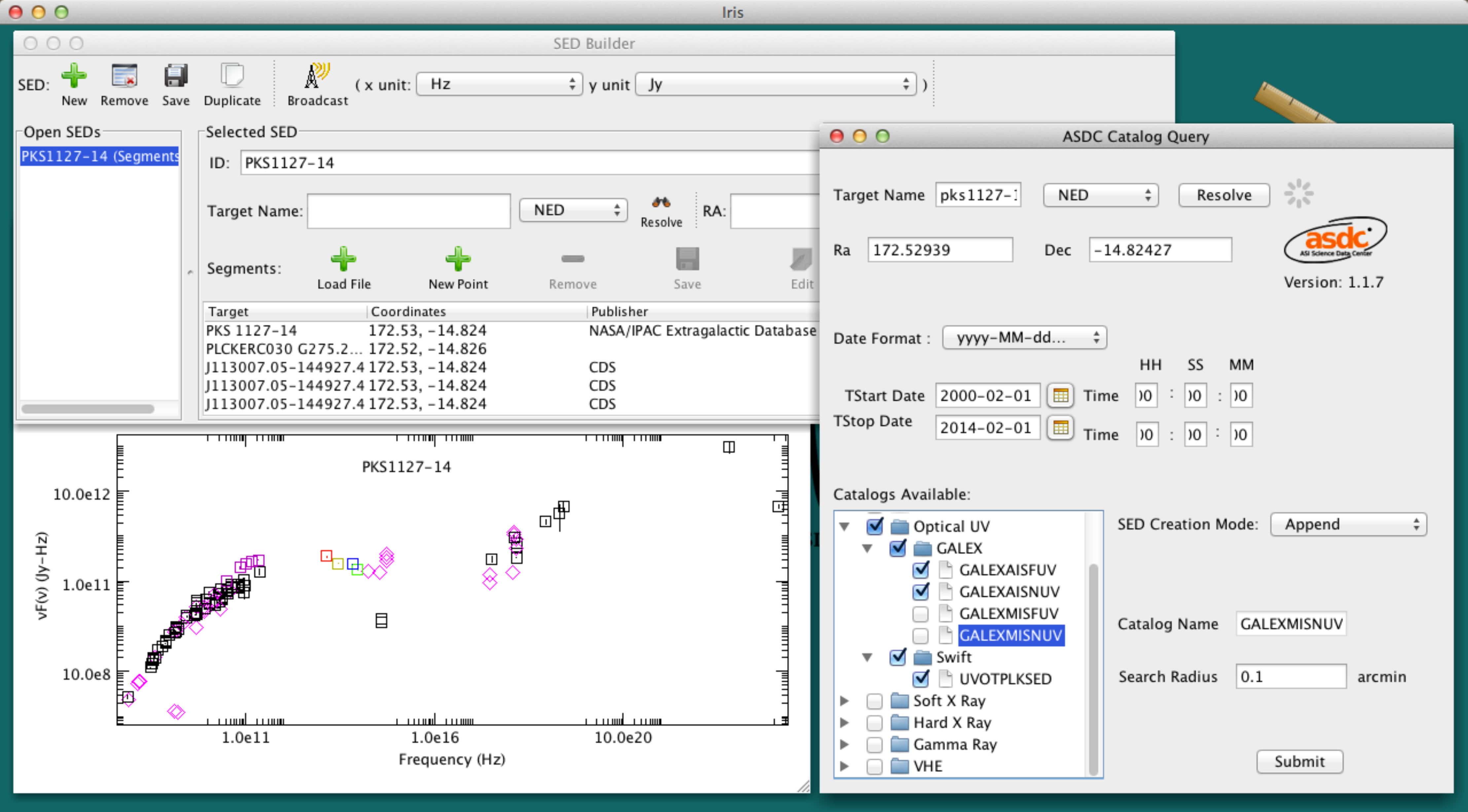}
\caption{\textbf{Building the SED of blazar PKS 1127-14 in Iris.}  
\textit{Top-left:} Data from the NED SED Service, a local file, and from TOPCAT
are managed in the SED Builder. \textit{Bottom-left:} The various data segments
plotted in $\mathrm{\nu F \left( \nu \right)}$ units inside the SED Viewer.
Squares show data with flux uncertainties, whereas the pink diamonds denote
points without associated uncertainties. Each segment in the SED Builder is
plotted in a different color. Black squares are data taken from NED; the
pink squares in the radio are the data from PLANCK; and the red, yellow, blue,
and green squares in the near-IR are the four WISE bands. \textit{Right:} An ASDC
Data Query form for PKS 1127-14. The user searches for data between specified
dates and available instruments (Swift and GALEX in this case). The data have
been added to the open SED PKS 1127-14.} \label{fig:load_data}
\end{figure*}

\begin{figure*} \centering
\includegraphics[height=2.95in,width=5.3in]{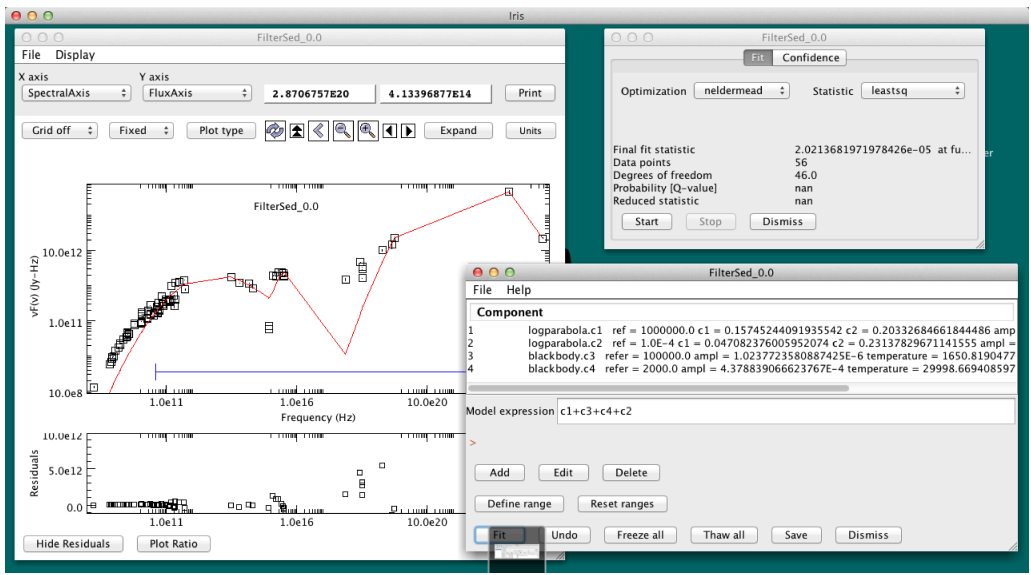}
\caption{\textbf{Fitting Visualization.}  Visualization of a linear combination of
log-parabolas and blackbody distributions for FRSQ blazar PKS 1127-145, fit with
Nelder-Mead optimization and least-square statistics. \textit{Left:} The best fit
linear combination overlaid on the SED data as a red curve. The blue line shows
the spectral range over which the data were fit. Below the main plot are
residuals of the fitted curve, in dex units. \textit{Top-right:} The fitting
options and results. Here, the user chooses between Nelder-Mead,
Levenberg-Marquardt, and Monte-Carlo \citep[Differential Evolution, ][]{Storn:1997:DEN:596061.596146}
optimization and various least square and
$\mathrm{\chi}^{2}$ statistics. The fit statistics are reported here after the
fit has been performed. \textit{Bottom-right:} The Fitting Tool window. The
model components used in the fit and their fitted parameter values are listed in
the Components field. Below that is the Model Expression, in which the
components are linearly combined. Note that components are referenced by the
\textit{c\#} suffix of the component name.} \label{fig:fitting1} \end{figure*}

\begin{sloppypar}
An Iris session begins with populating a SED by clicking on the SED Builder icon on the
Iris desktop. A user loads a local ASCII file of
PLANCK data, a WISE dataset from TOPCAT \citep[ascl:1101.010\footnote{ascl: Astronomy Source Code Library}]{2005ASPC..347...29T} 
through a SAMP message, and all data associated with PKS 1127-14 in NED with the NED SED
Service portal. The user also uses the built-in Italian Space Agency Science Data
Center\footnote{\url{http://www.asdc.asi.it/}} (ASDC) query tool to find
optical/UV data for PKS 1127-14, and adds it to the SED (see
Figure~\ref{fig:load_data}).
\end{sloppypar}

Data are converted to a single set of units on the fly, and displayed in the SED
Viewer. The user can switch the spectral and flux axes between a variety of
commonly-used SED units, e.g., one can switch from $\mathrm{Jy}$ vs.
${\mu}\mathrm{m}$ to $\mathrm{Jy}\mathrm{Hz}$ vs. $\mathrm{Hz}$. The Metadata
Browser --- an interactive table of the SED data --- allows the user to
interactively inspect and filter out data points by hand or with Boolean
expressions.

\begin{sloppypar}
The user also employs the Science Tools, an Iris built-in component that
lets the user cosmologically redshift SEDs, interpolate SED data, and calculate 
integrated fluxes of SEDs
through photometric filters or user-defined passbands. In particular, the user
shifts PKS 1127-14 from its observed redshift at $z=1.18$ to rest frame using
the Redshift tool before fitting the SED.
\end{sloppypar}

The user then filters out all the points devoid of errors using the metadata browser
filtering features.

When the user is done building and editing the SED, the user begins the fitting
session. With the fitting tool, the user can build a model expression as an
arbitrary combination of model components. Choosing from a list of built-in astrophysical
and mathematical models, the user fits PKS 1127-14 with a linear
combination of four models: two logarithmic parabolas to model the radio
synchrotron and inverse Compton radiation~\citep{2006A&A...448..861M,2009A&A...501..879T}, 
and two blackbodies to approximate the models for the hot dust component and 
accretion disk of the blazar~\citep{2002ApJ...575..667D}. The fit
is performed using Nelder-Mead optimization and least square statistics. The user
has fine control over the parameters, including setting initial values, the
range of the values, freezing and thawing parameters, and linking model
parameters to other parameters in the model expression; the user also controls the 
spectral ranges over which to fit the models. Finally, confidence intervals are computed 
for the overall model parameters.

Figure~\ref{fig:fitting1} shows the final model for PKS 1127-14 overlaid on the
input data and, in the lower panel, the fit residuals.

When the user is satisfied with the fitting results, the user saves an XML-style file
of the model that can be re-read into Iris and fit to other SED data. The user also
saves the fit results to a text file, that shows the parameters of the fit and
the details about each model component, with the best-fit parameter values.


\section{User Models and Templates} \label{sec:usermodels}

Keeping with our requirements of developing an extensible SED analysis tool, we
provide a user interface for adding custom models, templates, and template
libraries for the fitting engine to use in a Custom Fit Models Manager.

\begin{sloppypar}
Sherpa, Iris' fitting engine, provides command line functions for users to add
their own models and templates to a Sherpa session. We wrap a graphical user 
interface (GUI) around such functions for streamlined integration and 
user-friendliness.
The user provides the full path to the directory where the models and templates
exist, as well as information about the parameters. Installing a model saves a
copy of the model files in the user's home directory (in
\texttt{\~{}/.vao/iris/}components), allowing the user to apply the models in future
sessions.
\end{sloppypar}

\subsection{Custom Python Functions} Iris accepts custom models as Python
functions stored on the user's disk. Any number of functions can be stored in a
single file. The function implementing the model must take two parameters: the
first is an iterable of the model parameters, the second is a placeholder for
the spectral axis, $x$, in units of Angstroms. For example, a model file for a
modified black body \(B_{\nu}(T) \left(\nu/\nu_{0}\right)^{\beta}\) could be
defined as in Listing \ref{lst:user_model_example}.

User models can be arbitrarily combined with other custom or preset model
functions when using the Iris fitting tool.

\begin{lstlisting}[style=python,
	caption={Example of a user-defined model that
can be dynamically loaded into Iris. The code, written as a Python function, 
implements a modified blackbody and can be combined 
in Iris with other built-in and custom components. Backslashes indicate line 
continuations.},
	label=lst:user_model_example]
import numpy as np

def modified_blackbody(p, x):
  """ Modified blackbody.

  Parameters
  ----------
  p : [lambda_0, A, T, beta]
    p[0] 'lambda_0' : reference wavelength
    p[1] 'A' : amplitude of model at lambda_0
    p[2] 'T' : temperature of blackbody
    p[3] 'beta' : dust emissivity index
  x : array spectral values, in Angstroms
  """

  # Blackbody function
  efactor = 1.438786e8 / p[2]
  numerator = p[1] * np.power(p[0], 5.0) * \
              (np.exp(efactor / p[0]) - 1.0)
  denominator = np.power(x, 5.0) * \
                (np.exp(efactor/ x) - 1.0)
  B_lambda = numerator / denominator

  # speed of light in AA/s
  c = 2.998e18

  powerlaw = (c / (x/p[0]))**p[3]

  return B_lambda * powerlaw
\end{lstlisting}

\subsection{Table Models} A table model is a single template, having just the
$x$ and $y$ coordinates. Iris accepts two column ASCII files as table models,
following the convention where the first column is the spectral values
and the second contains the fluxes. The spectral and flux units must be in Angstroms
and $\mathrm{photons}/\mathrm{s}/\mathrm{cm}^{2}/\mbox{\AA}$, respectively\footnote{
While Iris ingests many other units (see Table~\ref{table:units}), the Custom Fit Model
Manager is independent of Iris's units handler and only accepts files with spectral 
values in Angstroms and the flux in 
$\mathrm{photons}/\mathrm{s}/\mathrm{cm}^{2}/\mbox{\AA}$}.
The fit returns the normalization constant (or amplitude) of the model.

\begin{sloppypar}
\subsection{Template Libraries} The template model is essentially a list of
table models with parameters other than the amplitude. Like the
\texttt{load\_template\_model} function in Sherpa, the user must create an index
file that lists the parameter values of the templates and the full path to the
template those parameter values describe (see Listing \ref{lst:templateconfig}
for an example). Sherpa uses a grid-search method to find the best-fit template.
The parameters grid is created using the values provided in the index file.
\end{sloppypar}

\begin{lstlisting}[style=code,
	caption={Example of template library definition file. Template library definition
	 files are in ASCII format.},
	label=lst:templateconfig]
# INDEX REFER MODELFLAG FILENAME
0.0     5000  1	/data/sed_temp_index-0.00.dat
-0.10   5000  1 /data/sed_temp_index-0.10.dat
-0.25   5000  1 /data/sed_temp_index-0.25.dat
-0.35   5000  1 /data/sed_temp_index-0.35.dat
-0.50   5000  1 /data/sed_temp_index-0.50.dat
\end{lstlisting}

\section{The Iris Stack} \label{sec:stack}

\begin{figure}
\begin{center}
\includegraphics[width=1.0\columnwidth]{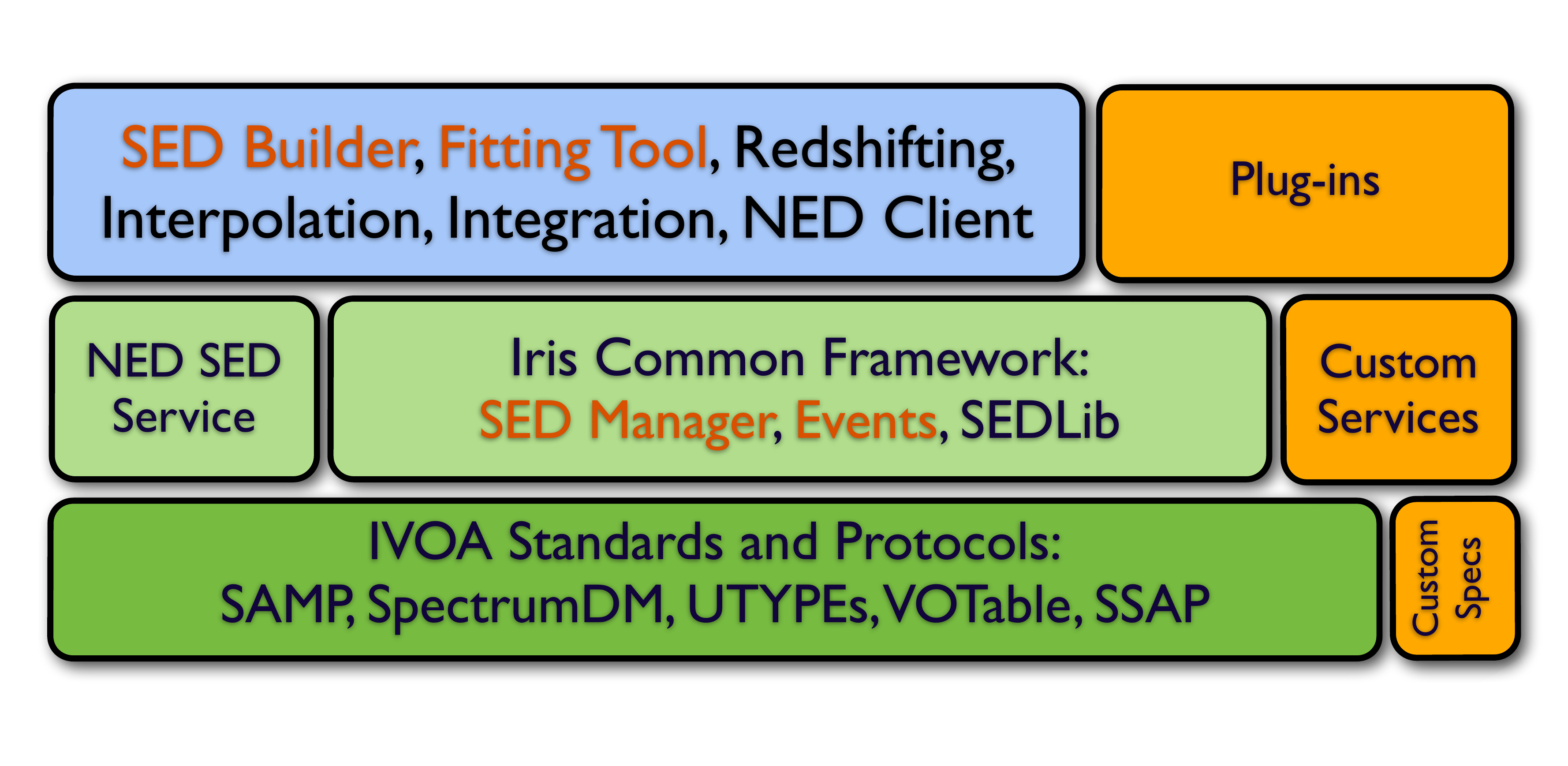} \caption{\textbf{The
Iris Stack.}  With its architecture Iris allows developers to create components
using higher and higher abstractions on top of web services, desktop
applications, and Virtual Observatory standards and protocols.
The technical
specifications lie on the bottom, a middle layer provides abstractions useful to
developers, and the user is only exposed to the science features represented by the top layer.
Users can also
plug their code in as Python functions. The result is aimed to be both user- and
developer- friendly. Notice that custom services can be built on top of custom specifications,
but also on top of IVOA standards. Similarly, plug-ins can use custom services, but they also
probably use the Iris Common Framework. The top layer components
(built-in Iris features and plug-ins) provide the user with scientific features within Iris.}
\label{fig:stack}
\end{center}
\end{figure}

\begin{sloppypar}
The Iris stack (Figure \ref{fig:stack}) shows how one can put the 
technical IVOA specifications to work for scientists through higher and higher
layers of abstraction: the details of the Virtual Observatory
standards and protocols lie in the lowest layer, the internals of the Iris
building blocks lie in the middle layer, while the top layer expresses
high-level user-oriented features.
\end{sloppypar}

A reader without any knowledge of programming, let alone of the VO
specifications, should understand the labels used in the top layer of the
diagram and their components (e.g., \textit{Fitting Tool} and \textit{Redshifting}), as long as they
have some knowledge of astronomical SEDs. On the other hand, a developer would
find words like framework, service, and manager quite familiar, while it takes a
VO-savvy person to decode the acronyms at the bottom of the diagram.\footnote{SAMP,
the Simple Application Messaging protocol was already introduced, DM stands for Data Model and the SpectrumDM
is introduced later in the paper. UTYPEs are labels used in some file formats (like VOTable, introduced later)
to tag data elements according to a Data Model. SSAP stands for Simple Spectral Access Protocol and
is implemented by services that provide access to spectral datasets, including SEDs, and is also introduced later
in the document.}

This architecture enables different entry points for the different audiences of the
application. Core developers work at all levels of the stack, but need to lay
out the foundations on top of the standard specifications; third party
developers use the middle-level abstractions offered by the Iris framework,
while end users can limit their interaction to familiar astronomical concepts
through the application's user interface. End users can also plug in their
modeling code and upload templates libraries to Iris.

The color code in Figure \ref{fig:stack} adds a different dimension to this
diagram and taps into a different characteristic of the Iris architecture:
extensibility. In particular, scarlet letters denote extensible
components of the architecture, i.e., components that offer hooks into the Iris
architecture to users and developers. The orange boxes, on the other hand,
express components that were not part of the Iris design, but that can be
used in Iris as plug-ins, possibly providing interfaces to access
non-standard services. Some of these plug-ins, along with a description of the
design of the Iris Software Development Kit, will be introduced in Section
\ref{sec:plugins}.

The dark green box denotes IVOA sanctioned standards. Blue denotes components
that are built-in in Iris and light green boxes denote components that were developed
in or for Iris.\footnote{While the NED SED service was developed independently of Iris,
its IVOA-compliant interface was part of the Iris project, along with the development of a
dedicated client in Iris itself.}

This architecture was also driven by a more abstract requirement:
our team was
distributed, with developers and managers working from different
institutions with different tools and practices \citep{2012SPIE.8449E..0IE}.
Moreover, wanting to reuse existing software instead of reinventing the
proverbial wheel, we had to integrate different existing software components in
a seamless way. So, the Iris stack provided not only a clean and robust architecture
for users and third party developers, but was also useful in enabling a distributed
team of part-time developers to work in parallel, reducing the overall project risk.

In summary, the Iris framework was designed to address several different
requirements: (i) functional requirements gathered by the Iris team's lead
scientists; (ii) functional requirements unknown at development-time; (iii) the
distributed nature of the Iris development team; and (iv) interoperability between 
several existing tools and services.

The Iris stack offers a non-technical view of the Iris architecture and design.
While the stack shows effectively how we tried to abstract end users and developers
from the VO specifications and from the specifics of the Iris internals, the stack does
not express the technical solutions that we employed to achieve such extensible
architecture and to meet the aforementioned requirements. More detail is
provided in some of the following sections.

\section{The Iris Architecture} \label{sec:architecture}

\begin{figure*} \begin{center}
\includegraphics[width=0.7\textwidth]{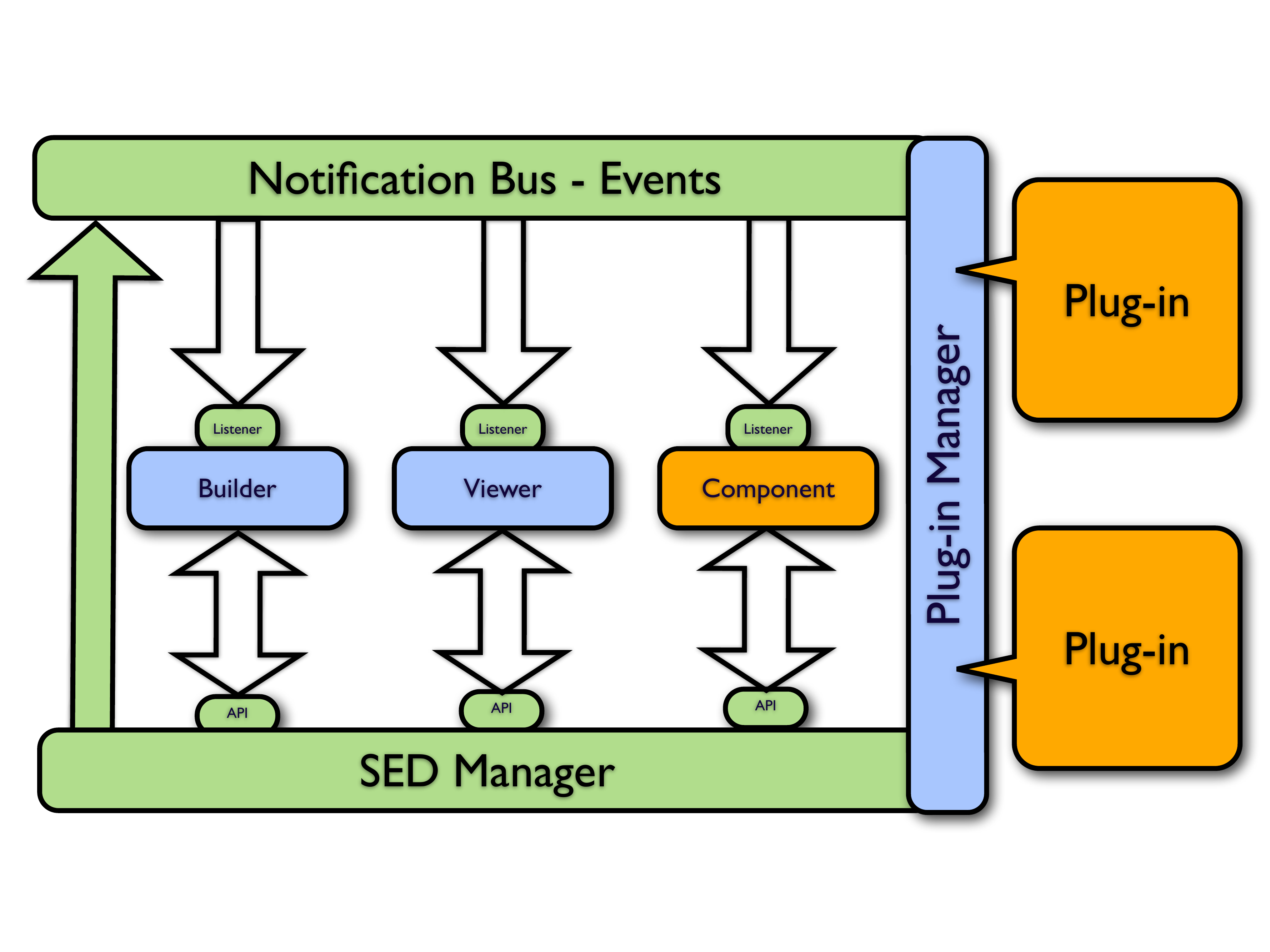}
\caption{\textbf{The Iris loosely coupled, extensible architecture.}  Information
freely flows among built-in and third-party components provided as plug-ins. A
SED Manager gives components access to the state of the SEDs in the user's
workspace, while dynamic changes in such state are announced through events that
are notified to the subscribed listeners. A plug-in manager allows users to
install and uninstall plug-ins on the fly.} \label{fig:architecture}
\end{center} \end{figure*}

\begin{figure*} \begin{center}
\includegraphics[width=0.6\textwidth]{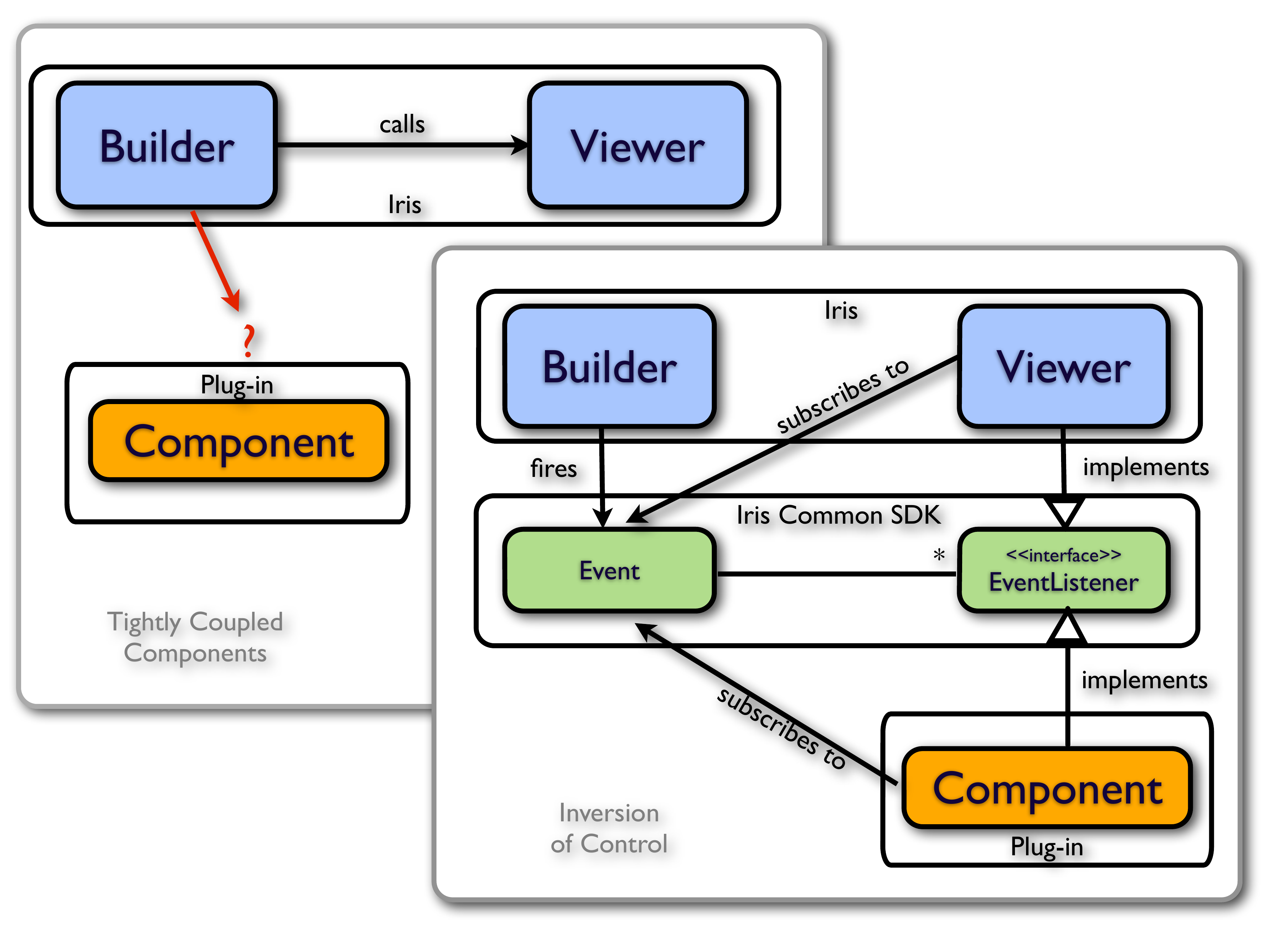}
\caption{\textbf{Inversion of Control.}  IoC is a design pattern that promotes
decoupling of software components so that they can easily be replaced by
different implementations, with the actual binding often happening at run-time
according to some configuration. This pattern, however, also allows new
components to be added at any time during the application lifecycle: a common
framework (Iris Common) can be shared as a middle layer among implementations,
and a container (the Iris application) can bind components together on the fly.
Components can subscribe to events and react to them.} \label{fig:ioc}
\end{center} \end{figure*}

In order to minimize the risk derived from the requirements listed in the previous 
section, we backed Iris with a loosely coupled architecture through a design 
pattern called Inversion of Control \citep*{ioc}. 

But it was not just a matter of risk management. Inversion of Control  
supports the implementation of \emph{liquid} requirements, i.e., a finite set of
predetermined requirements plus an indefinite set of custom requirements to be
implemented by users, at least in some simple cases or, for more advanced
features, by third party developers.

The architecture that supports the implementation of such requirements has
different components that can be mapped to the Model-View-Controller (MVC)
design pattern.
\begin{description}
\item[SEDLib] This basic I/O library provides
classes for the Model components of MVC. Unsurprisingly, SEDLib does so by
implementing a Data Model specification defined by the IVOA. The Data Model
defines both the logical breakdown of spectral datasets, and the serialization
in some standard file formats. So, on the one hand, SEDLib can perform the basic
read/write operations on spectrophotometric files, while on the other the library
provides the data structures that client components can use and exchange.
\item[SEDManager] The MVC Controller role is played in Iris by the SEDManager,
which itself is defined as an Interface. The manager works as a data storage for
SEDLib instances that the different Iris components can share.
\item[Components] The actual Iris functionality is im\-ple\-ment\-ed by the Iris
Components. They can be seen as the Views in the MVC pattern (or, more
generally, they can provide any number of Views), since they present the data
stored in the Controller to the user, query the Controller itself, and act upon
the Models, i.e., the SED objects provided by SEDLib. 
\item[Events] Views can be
notified of changes in the Models by Events, if they implement the relative
Listener interface and have been registered to the Events Queue. Events usually
have a payload with more information about their content, and a pointer to the
Model instances involved.
\end{description}

\begin{sloppypar}
In summary, Components (Views) can be completely disentangled from each other
and interact indirectly through the sole common interface represented by the
SEDManager (Controller), which in turn stores the SED objects (Model). Dynamic
changes in the system are notified to all interested agents (Listeners) via
specific Events.
\end{sloppypar}

Components are thus agents that cooperate by attaching themselves to a common
\emph{bus} where the SEDManager provides the memory, and Events guarantee the
flow of information (see Figure~\ref{fig:architecture}).

\subsection{Inversion of Control}

We achieve loose coupling by an extensive use of Java Interfaces: components,
events, and event listeners, for example, are all defined by interfaces whose
implementation can, to some extent, be freely interchangeable.

Moreover, Inversion of Control is employed to decouple
the implementation of components from the run time context (see Figure~\ref{fig:ioc}). 
Methods in the
Interface are callbacks, and some of these callbacks get Interface-typed
arguments that provide them context instances during application execution.
For this reason, this pattern is also sometimes referred to as \emph{Dependency
Injection}\footnote{There is, to be precise, a subtle but significant difference
between Dependency Injection and Inversion of Control, the first effectively
being a special case of the second.}.

Consider, for example, Iris Components: they are the main providers of Iris
functionality, and they can correspond to buttons and menu items on the Iris
desktop, loggers, data handlers, etc. They must implement the \texttt{IrisComponent}
interface, listed in Listing \ref{lst:component}.

\begin{lstlisting}[style=java,
	caption={This snippet of Java code represents the main interface that all components in Iris have to implement, and how dependencies get injected into the components at run-time. Backslashes indicate line continuations.}, 		
	label=lst:component]
package cfa.vo.iris;

import java.util.List;
import org.astrogrid.samp.client.MessageHandler;

public interface IrisComponent {

  /**
   * This method is invoked to initialize the
   * component. If the component has to
   * launch windows, frames or background
   * services, this is the right method to do
   * so. Otherwise the component will be
   * called only if a callback is invoked.
   * @param app A reference to the running application
   * @param workspace A reference to the application's workspace
   */
  void init(IrisApplication app, IWorkspace workspace);

  /**
   * Return the name of this component. This
   * name might be listed in a widget along
   * with the other registered components.
   * @return The component's name as a String. 
   */
  String getName();

  /**
   * Get the description for this component. The
   * description might be listed in a widget
   * along with the other registered components.
   * @return The component's description as a
   * String.
   */
  String getDescription();

  /**
   * Get a command line interface object for
   * this component.
   * @return A CLI object
   */
  ICommandLineInterface getCli();

  /**
   * Initialize the Command Line Application
   * interface
   * @param app Reference to the enclosing
   * application
   */
  void initCli(IrisApplication app);

  /**
   * The component can contribute menu items
   * and desktop buttons to the enclosing GUI
   * applications by providing a list of
   * MenuItems.
   * @return A list of the menu items this
   * component will contribute to the
   * application. 
   */
  List<IMenuItem> getMenus();

  /**
   * The component can register any number of
   * SAMP message listeners by providing a
   * list of them.
   * @return A list of the SAMP message
   * listeners that have to be registered to
   * the SAMP hub. 
   */
  List<MessageHandler> getSampHandlers();

  /**
   * Callback invoked when the component is
   * shutdown
   */
   void shutdown();
}
\end{lstlisting}

At startup the Iris application reads the list of Components to be initiated,
and calls their \texttt{init} call-back, which in turn is passed useful information
like a reference to the SEDManager, or hooks to the application environment.

The advantages of this architecture are both functional and non functional. The architecture
helped our heterogeneous development team to work in a loosely coupled way,
reducing the overall project risk, and also provided the extensible framework
we were seeking in the first place. As a matter of fact, plug-ins that can be
loaded at run time implement the same interfaces that the built-in components
do, and they are instantiated in exactly the same way. The only difference is in
the timing: built-in Components get instantiated when the application itself is
initialized, while plug-ins can be instantiated and discarded at any time during
the application execution.

\section{Iris Built-in Components} \label{sec:components} In the previous
section, we discussed the architecture of Iris and how the different Components
in Iris communicate. Each Component performs one or more SED-related tasks in
Iris, like building SEDs from multiple sources and fine-tuned SED modeling.
Here, we discuss what the Components do in terms of the science domain, including
descriptions of the autonomous software used to build
Iris: Specview, Sherpa, and the NED SED Service.

\begin{figure*} \begin{center}
\includegraphics[width=0.9\textwidth]{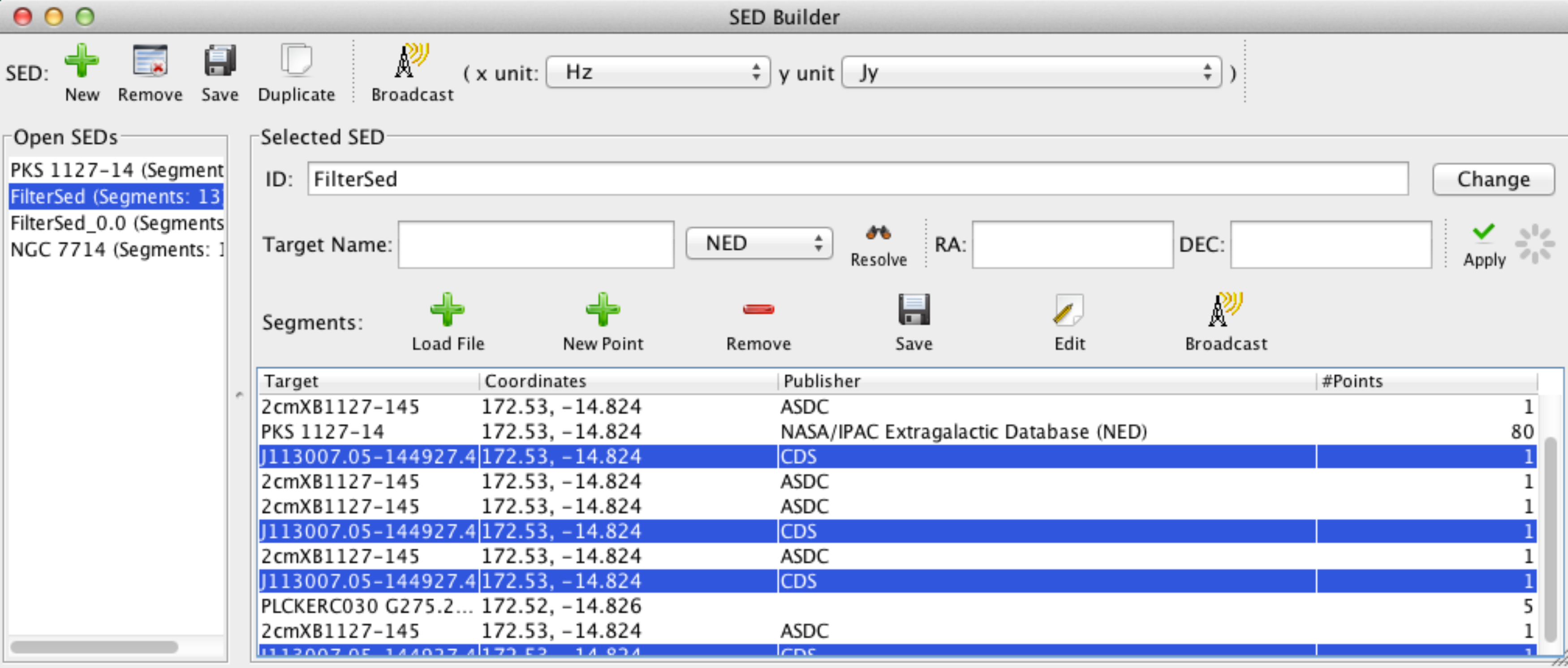}
\caption{\textbf{SED Builder.}  An example view of the SED Builder. On
the left side of the window is a list of SEDs open for analysis; in this case,
\textit{FilterSed} is selected. The Segments or components that constitute
\textit{FilterSed} are shown in the Segments section. Segments may be managed
separately. Highlighted SEDs or Segments may be edited, removed, saved, or
broadcast to an external SAMP-enabled application. New Segments and SEDs are
added to the Iris session through the SED Builder. } \label{fig:sed_builder}
\end{center} \end{figure*}

\subsection{SED Builder}

Users manage SEDs through the SED Builder (Figure~\ref{fig:sed_builder}). From
the Builder, users can add, edit, remove, and save SEDs. Users can also transfer
data seamlessly to other VO-enabled applications through SAMP messages from the
Builder. Any number of SEDs can be analyzed in an Iris session. Each SED has a
unique identifier that is set by default when a new SED is created, but can be
changed by the user. The user switches between SEDs by clicking on a SED name in
the Open SEDs field; the visualizer will automatically update to the selected
SED.

SEDs are built and managed in Segments, which are groups of (spectral, flux)
coordinates. For example, a spectrum is considered a Segment; the results of a
NED SED Service query are also handled as a Segment. In general, anything from a
single photometric point to an entire SED can be considered a Segment, with all the
points sharing some if not all of the metadata.

Clicking on a SED in the Open SEDs field will show all the Segments that
populate that particular SED. SED Builder shows where the Segment data came
from, the recorded RA and Dec of the Segment, and the number of points in the
Segment. Segments can be handled separately from other Segments in the SED;
users can add, edit, remove, and save a subset of Segments selected from a SED.

\subsubsection{Importing data}

\begin{figure}[h!] \begin{center}
\includegraphics[width=\columnwidth]{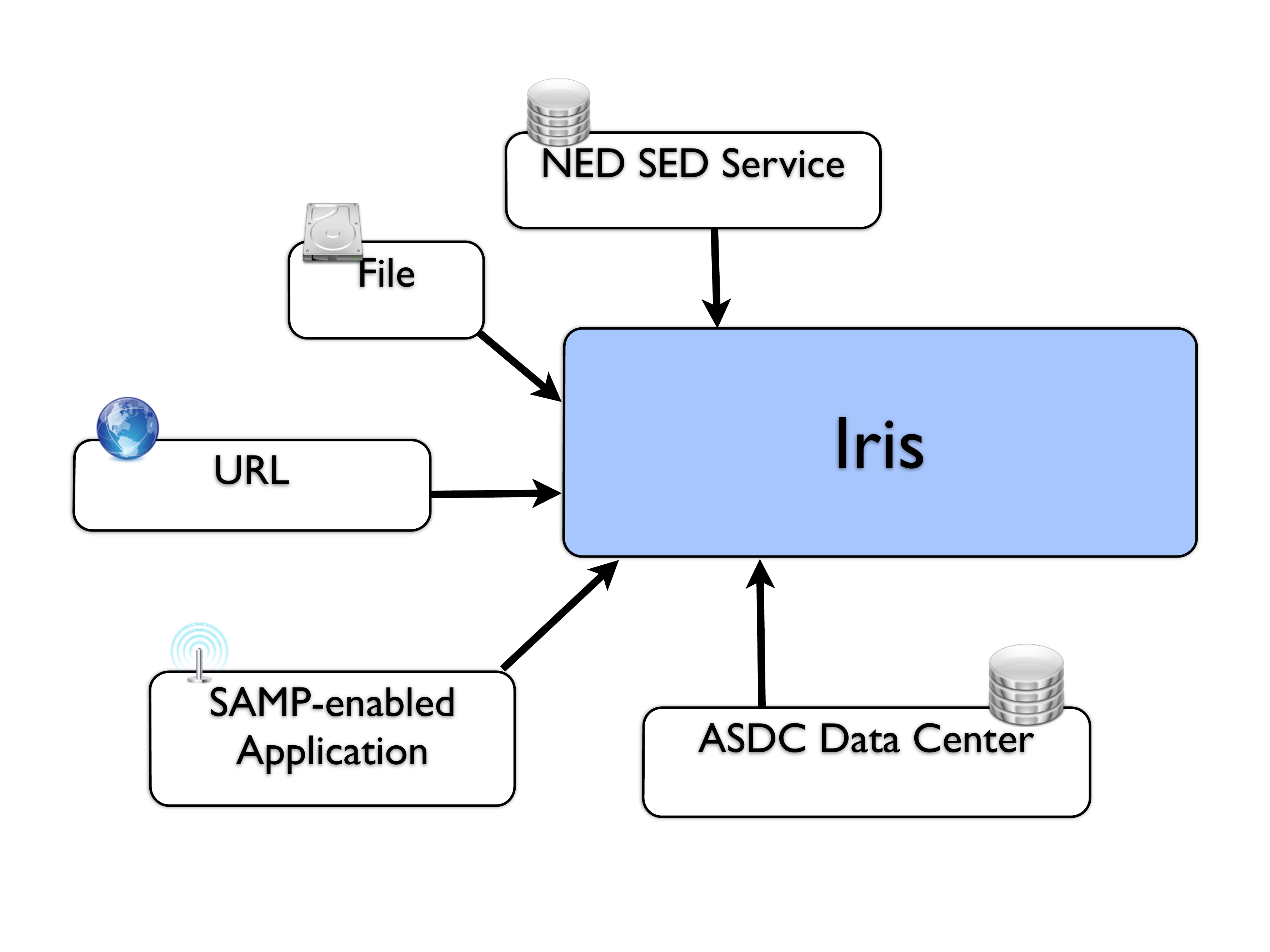}
\caption{\textbf{\label{fig:data_sources} Available Data
Sources.}  Users may import data from the built-in
clients to the NED SED Service and the ASDC. Data may also be uploaded from 
a local file, a URL, a SAMP message from a VO-enabled tool, 
or through a custom file filter plug-in at run time.} \end{center}
\end{figure}

As described in Section~\ref{sec:overview}, Iris accepts data from a variety of
sources, and is lenient on the data format. Figure~\ref{fig:data_sources}
illustrates that Iris imports data from built-in data archive portals as well as
from outside resources like local files, URLs, other VO-enabled applications,
and from plug-ins.

Iris natively supports IVOA-compliant FITS and VOTable formats
\citep{2012arXiv1204.3055M}. Files in these formats will automatically be added
to the user's workspace. The Builder can convert ASCII Tables, CSV,
TSV, IPAC tables, and non IVOA-com\-pli\-ant VOTable and FITS files
into the native format with user input. We provide two importing forms: (i) the
SED importer, which handles spectrum-style files (i.e., those with columns for
the spectral coordinate, flux/energy, and flux/energy uncertainties), and (ii)
the Photometry Catalog Importer, which handles photometry catalogs (i.e., files
where each column represents a passband and the cell values represent the
corresponding fluxes, with an arbitrary number of rows). Users can save their
setup options from the Import Setup Frame to a configuration file and
automatically read-in files of the same format to Iris via the command line.

The SED Builder also has a hook for adding custom file filters. One could develop
a custom file reader that would convert a non-standard file to an IVOA-compliant
format. This kind of add-on would allow Iris to read non-standard
files into Iris without requiring the use of the importer tools.

\subsubsection{Saving data} Users can save entire SEDs or sets of Segments to
IVOA-compliant VOTable or FITS files. In order to save all the metadata, the
IVOA-compliant serializations rely on some specific constructs in the supported
file formats, so that SEDs that have many different Segments can become very
complicated to read for VO-unaware applications, although they retain all the
metadata details. For instance, segments might have data expressed in different
units inside the same SED.

To facilitate the ingestion of SEDs in VO-unaware applications and user scripts,
we provide a simpler output format that only saves the minimum amount of
meaningful information: the spectral coordinate, the flux or energy, and its
uncertainties. As a result, the resulting SED file has only
one Segment, with all the data expressed in a single set of units defined by the user.

Whether the output includes all of the metadata or has a simplified \emph{single
table} format, the result is a compliant file that can be read back into Iris
without any additional user's input.

This allows users to save a standardized version of the file that can be easily
shared by Iris and by the user's scripts.

\subsection{NED SED Service} \label{subsec:ned}

Iris is packaged with a portal to the NED SED
Service that, given a target name,
retrieves all photometric data in NED associated with the source with that target
name, and adds it to an existing SED.

In the context of the VAO development of Iris, we adapted the NED long standing photometry and spectral energy
distribution service to conform as closely as practical to the relevant
IVOA recommendations in order to deliver photometric data from the
collection into Iris seamlessly. The objective for NED was to provide a working reference
service for the development of Iris as well as to serve as a working prototype
for new data protocols for spectrophotometric data being developed by the IVOA.

\begin{sloppypar}
The NED SED Service returns data and information from the NED photometry
collection \citep{2007ASPC..376..153M}. The NED SED Service provides three types
of queries:
\end{sloppypar}

\begin{description}
\item[Information Discovery] List objects with
available photometry (SED) given a sky position (RA and Dec) and angular size.
Also called a \emph{data discovery} query. 
\item[Information Availability] For
a given named object, return the number of photometric data points. 
\item[Data Retrieval] For a given named object, return the
available photometric data in an IVOA Spectrum Data Model compatible VOTable.
\end{description}

All three query types use HTTP requests and responses which conform to the 
IVOA Simple Spectral Access Protocol Version 1.04 \citep[SSAP;][]{2012arXiv1203.5725T}; 
the responses are in VOTable format. The NED SED Service client in Iris
employs the Data Retrieval query interface, and stores the
response as a Segment. Photometric points with spectral line-based values and
upper- and lower-limit values are excluded from the response.

Implementing a standard protocol interface, the NED SED service is also
available through generic VO applications like TOPCAT and the VAO Data Discovery
Tool\footnote{\url{http://vao.stsci.edu/portal/Mashup/Clients/Portal/DataDiscovery.html}}.

\subsection{SED Viewer} \label{subsec:specview}
The Iris Viewer component is responsible for creating, managing, and providing
user interactive feedback to spectral plots in Iris.

The Viewer also provides most of the low-level GUI
components used by the Fitting Tool component. The reason for this is that most,
if not all of the GUI code used by both the Viewer and the Fitting Tool, were
developed on top of the Specview \citep[][ascl:1210.016]{2002ASPC..281..120B} code base.

Specview was developed in the late 1990's, initially as an experiment to evaluate
Java graphics capabilities in the context of interactive spectral plotting. Over
the years Specview grew from a simple visualizer dedicated mostly to plot
spectral data from Hubble Space Telescope (HST) instruments, to a more capable
tool with not only sophisticated visualization, but also data analysis
capabilities. The ability to ingest spectral data from a variety of sources was
also gradually incorporated into the tool, culminating with a Virtual
Observatory interface capable of accessing services that comply with the
SSAP standard.

Specview however kept the emphasis on spectral data, which is very different from
the broad-band SED concept to which Iris is dedicated. Being initially conceived
as a tool to support HST data, the design, and subsequent code implementation,
were driven by the needs and requirements imposed by high-dispersion, relatively
narrow-band spectra in the near-IR / optical / near-UV range. Thus some re-work
was necessary to make Specview's internal data structures and algorithms comply with
the data types associated with SEDs. Even so, a significant part of the code
could be kept as is, thus realizing the savings associated with code re-use.
This is particularly true in the case of the low-level graphics
engine~\citep{2000ASPC..216...79B}. Most of the work in adapting Specview's code
base to Iris happened on two fronts: (i) adding code that implements the Iris
Component interface, and (ii) augmenting the capabilities of the Data Browser to
allow interactive access to SED metadata. Some work was also done in fine-tuning
plotting capabilities to the particular needs of SED data.

The initial view the Viewer creates of a just-ingested SED is via a scatter plot
depicting wavelengths (frequency and energy units are also supported) and flux
density (or flux) for each data point that comprises the SED. The plot can be
configured in a variety of ways, by changing the scaling and units. The data
initially plotted can then be further examined in more detail, using tabular and
tree depictions. In particular, the metadata associated with each data point, as
well as the global metadata associated with the entire SED, can be examined in
detail using the Metadata Browser. Data points can be selectively removed from
the SED using filters sensitive to both data and metadata values. These filters
are built by a user-defined Boolean expression that can be created and
interacted with in the GUI itself. The expression uses Python-like syntax, and
Python operators are available throughout. That way, SEDs can be modified after
being read by the SED Builder, and before being further processed or measured.

\subsection{Sherpa: Model Fitting}
\label{subsec:sherpa} 

\begin{sloppypar}
Sherpa (ascl:1107.005) is the Chandra Interactive Analysis of Observations
\citep[CIAO;][ascl:1311.006]{2006SPIE.6270E..60F}
modeling and fitting application. Sherpa enables the user to construct complex
models from simple definitions and fit those models to 1D (spectra) and 2D
(images) data using a variety of statistics and optimization methods.
\end{sloppypar}

Written in Python, with C/\Cpp/Fortran extensions, Sherpa was a robust choice
for providing Iris with a curve fitting engine.

However, since the Iris front end was going to be a Java application\footnote{In
the first version of Iris the front end was a modified version of Specview
itself, while in later versions we integrated different components under a
common framework graphically represented by the Iris Desktop. Even in this
configuration, the fitting front end was provided by Specview under the hood.},
an interoperability layer had to be designed to interface the graphical user
interface and Sherpa as a fitting engine back-end.

SAMP is used as the interface protocol. This decision makes the design of the
interface very simple, so that the interoperability layer on top of Sherpa is
rather thin and consists only of the code required to inspect the incoming SAMP
messages and build a call to Sherpa.

\begin{figure*} \begin{center}
\includegraphics[width=\textwidth]{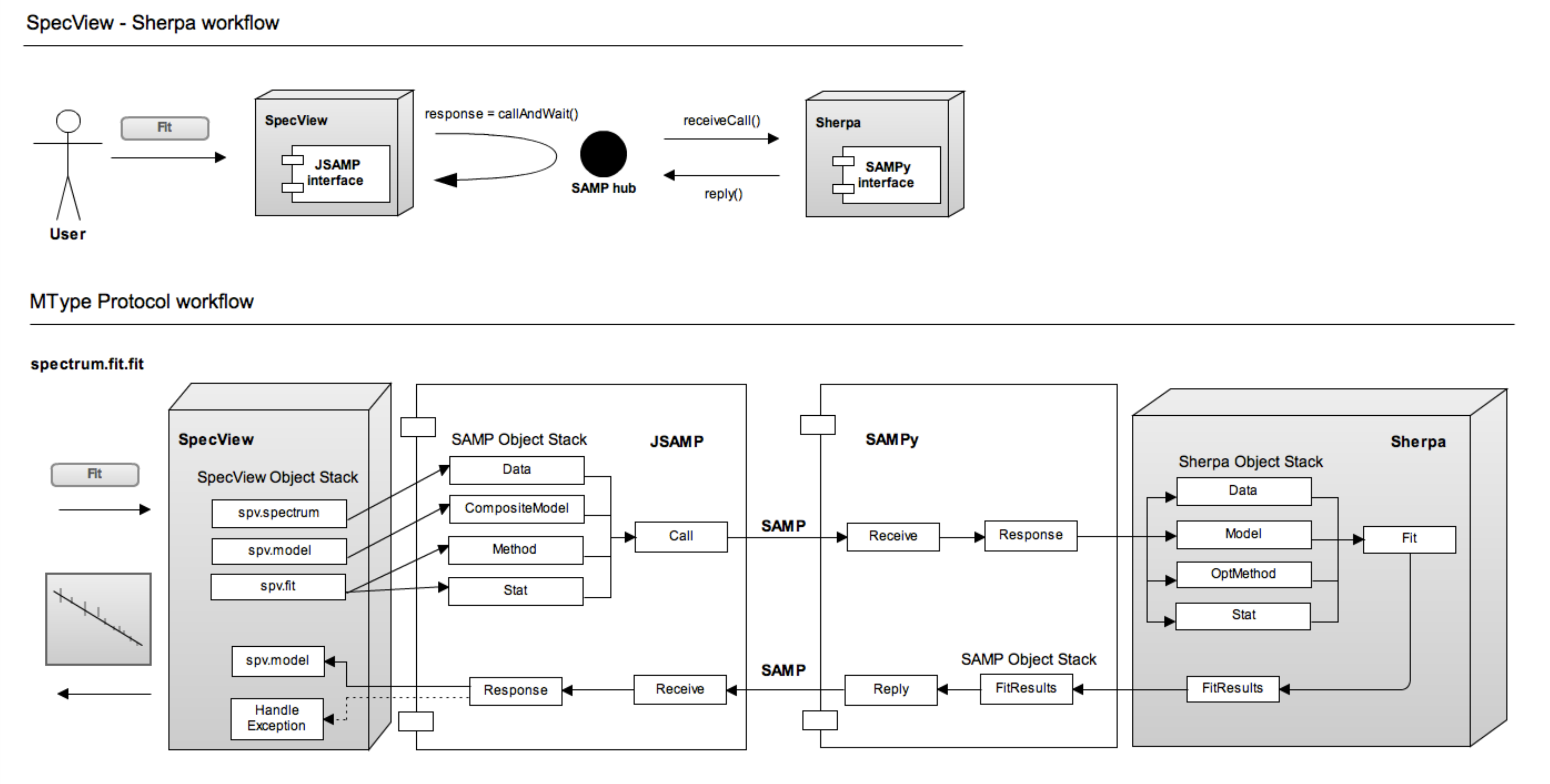}
\caption{\textbf{Design of the Specview/Sherpa interoperability layer.}  The
interface between Sherpa (fitting engine), and Specview (graphical user
interface) was designed by defining a common data model for representing the
requests and responses of fitting operations: the serialization of the request
is a SAMP message, whose \texttt{mtype} identifies the remote operation that the
client is requesting.} \label{fig:sherpasamp} \end{center} \end{figure*}

The design of this interface is represented schematically in Figure
\ref{fig:sherpasamp}.

When Iris is launched the \verb|sherpa-samp| process is also started in the
background. This process starts a SAMP client that waits for a SAMP hub to
attach to, registering to a number of custom
\verb|mtypes|. The \verb|mtypes| work as remote procedure identifiers, and SAMP
messages provide the remote methods with data that need to be processed. Sherpa
is used to compute a response that is packaged as a
SAMP response to be shown to the user.

\begin{sloppypar}
The thin layer between Java and Python code is implemented using two existing
implementations of the SAMP protocol, namely
\verb|jsamp|\footnote{\url{http://software.astrogrid.org/p/jsamp/1.3/}} for Java and
\verb|SAMPy|\footnote{\url{http://pythonhosted.org//sampy/}} for Python.
\end{sloppypar}

The \verb|sherpa-samp| layer grew to accommodate the new science requirements in
the latest Iris releases, so to include some analysis code that is
independent of Sherpa.

\subsubsection{Fitting Options}
\label{sherpa-fitting}

We provide the following fitting optimization methods and fit statistics from Sherpa 
in Iris. \citet{2009pysc.conf...51R} discuss Sherpa's statistics in detail. Here, 
we briefly present the options.

The optimization methods available in Iris are variations of the Nelder-Mead simplex, 
Levenberg-Marquardt, and Monte Carlo algorithms. The Nelder-Mead simplex method, which
finds the local minimum of a function in parameter space through a direct search method, is an
adaptation of the algorithms described in 
\citet{wright1996} and \citet{Lagarias:1998:CPN:588893.589108}. 
Levenberg-Marquardt optimization 
finds the local minimum of non-linear least squares functions of the model parameters
\citep{jjmore1978}. Lastly, the Monte Carlo method uses a differential evolution
algorithm outlined in \citet{Storn:1997:DEN:596061.596146} 
to find the \textit{global} minimum in parameter space.

Sherpa provides several $\chi^{2}$ statistics with different variances. For example, 
users can use the variance of the 
$y$-uncertainties (or $y$-values if there are no uncertainties), 
or they can set the variance to 1.
Also included are two maximum likelihood functions based on Poisson statistics: Cash and 
C-statistic \citep{1979ApJ...228..939C}. 



\subsection{Science Tools} We provide built-in science tools that perform
calculations commonly used in SED analysis: redshifting, interpolation, and
integration. The data are setup on the Java-side of Iris, but the actual
calculations are performed in \verb|sherpa-samp|.

The open SEDs are listed in the Science Tools frame. The user selects the SED
they wish to analyze, and inputs the required information for a calculation.

\subsubsection{Redshifting} Redshifting SEDs in Iris refers to cosmological
redshift. The spectral values are transformed into wavelength-space before 
shifting the SED.
Because the apparent magnitude of a source is dimmer at high redshifts
than low redshift, we correct the flux so that the area under the shifted SED
equals that of the un-shifted SED using
\begin{equation}
\label{eq:redshift} 
f_{z_{f}}(\lambda) = f_{z_{i}}(\lambda) \frac{\sum_{k=1}^N
(f_{z_{i}}(\lambda_{k+1})+f_{z_{i}}(\lambda_{k}))}{\sum_{k=1}^N
(f_{z_{f}}(\lambda_{k+1})+f_{z_{f}}(\lambda_{k}))}, 
\end{equation}
where $f_{z_i}$ is the observed flux at the initial (observed) redshift $z_i$,
$f_{z_f}$ is the flux at the final (target) redshift $z_f$, ${\lambda}$ is the
wavelength, $N$ is the number of points in the SED, and $\lambda_{k}$ is the wavelength
of the $k^{th}$ point in the SED. In \verb|sherpa-samp|, we extend the
astLib\footnote{\url{http://astlib.sourceforge.net/}} \texttt{astSED} class that
implements Equation~\ref{eq:redshift}.

From the the user's perspective, the user supplies the initial and final redshift of the
SED and clicks ``Create New SED.''

\subsubsection{Interpolation} Iris provides 1D interpolation along the spectral
axis. There are three interpolation options: linear, linear spline, and nearest
neighbor. Interpolation may be carried out on a linear or logarithmic scale.
Users may choose the number of bins, the spectral range over which to
interpolate, and may choose to smooth the resultant SED via a boxcar method.

\subsubsection{Integration}
The Integration tool was developed for
estimating integrated fluxes of a SED. The tool acts as a wrapper of the
\texttt{astLib} methods \texttt{calcFlux} and \texttt{integrate}, which in turn use
the composite trapezoidal rule to integrate the SEDs.

Iris provides two
methods of integration: (i) through a user-defined passband, and (ii) through a
photometric filter. The first option lets the user specify the spectral range in
wavelength, frequency, or energy units (Angstroms, Hz, and keV, respectively) to
integrate under. The second estimates the integrated flux measured through any
of the photometric filters provided by the Spanish Virtual Observatory's (SVO's)
Filter Profile
Service\footnote{\url{http://svo2.cab.inta-csic.es/theory/fps/index.php}}
\citep{2013arXiv1312.3249S}. This service has an extensive collection
of over 1000 filters at IR, optical, and UV
instruments. The user chooses from a list of filters
that can be searched by double-clicking on an instrument name, or by searching
for a string in the browser. The user sees the minimum, maximum, and effective
wavelengths of the filters before applying the filter to the SED.
Both methods return the effective wavelength of the passband in Angstroms and
the calculated flux in Jansky. The user can export the data to a new SED or
save the results to a simple ASCII formatted file.

\begin{sloppypar}
Notice that Iris currently integrates the SED data points, possibly after an interpolation,
and not the model. So, if the transmission curves or passbands do not completely overlap with the SED
Iris will return a NaN (Not a Number).
\end{sloppypar}

\section{Plug-ins: the Software Development Kit} \label{sec:plugins}

Iris offers a Java Software Development Kit (SDK) that can be used to extend the Iris
capabilities through the use of dynamically pluggable add-ons, or plug-ins.  The
use cases for this are listed below.
\begin{description} \item[New
functionality] A developer may want to add new capabilities in one or
more new Components. This use case can be broken down in more detailed and
concrete extensions, described later in this section.  \item[Custom-to-Standard
adapters] A developer may want to create adapters that query a non-standard
service, or load a non-standard dataset, and then turn the data to SEDLib
objects, thus effectively standardizing them so that they can be used by other
components in the Iris environment, or reused by other VO applications. In other
terms, one can achieve interoperability using the Iris infrastructure starting
from a non-interoperable service, file, or tool. Iris actually has some built-in
Custom-to-Standard adapters, like the \verb|sherpa-samp| layer described in Section
\ref{sec:components}, or the ASDC plug-in interface that queries a
quasi-standard service, described in Section \ref{sec:asdc}.  \end{description}

This section describes Java plug-ins, while Section \ref{sec:usermodels} described how
users can extend the models for fitting SEDs using Python functions.

\subsection{Anatomy of a Plug-in} A single Java Archive (jar) file can contain several
plug-ins, and each plug-in can bundle several Iris Components.

Each Component can provide several additions to Iris, as described in some
detail below.

\subsubsection{Menus and Buttons} Usually, although not always, an Iris
Component is visible to the user as either a set of buttons on the Iris Desktop,
or as a set of menu items in the Iris menu bar, or both.

Menu items can be added to either the File menu or to the Tools menu in a
specific plugin-related folder.

While the implementation of such buttons and menu items could be done from
scratch by implementing some Java Interfaces, a set of abstract classes
implements a lot of the boilerplate code and makes some convenient assumptions.
This way buttons and menu items can be created with very few lines of code.

Menu items and buttons can be customized by providing the button name, a
description that will be rendered as a mouse-hover tooltip, and icons.

\subsubsection{Command Line} Iris offers a framework for providing simple
command line interfaces to its tools. For example, Iris ships a command line
interface to the SED Builder (see Section \ref{sec:components}) that allows
users to import non-standard files in bulk through scripts, possibly starting
from templates saved interactively from the SED Builder.

The framework is extensible through a simple dispatching mechanism. Each
component has a name that is used to dispatch the command line argument to the
right CLI engine. For instance, the line
\begin{lstlisting}[style=code]
./Iris builder config.txt
\end{lstlisting}
instructs Iris to dispatch the \verb|config.txt| argument to the SED Builder's
CLI engine. Components bundled with plug-ins can provide such an engine by implementing the
\verb|ICommandLineInterface| Java Interface as shown in Listing \ref{lst:cli}.

\begin{lstlisting}[style=java,
	caption={Every Iris component can expose a command line interface. Iris dispatches the command line arguments for the relative component to process. This code is written in Java.},
	label=lst:cli]
package cfa.vo.iris;

/**
 * A simple interface for providing CLI access in 
 * an extensible, pluggable way
 * @author olaurino
 */
public interface ICommandLineInterface {

  /**
   * The name that has to be associated with the
   * implementing component.
   * When the calling application parses the 
   * command line, it will interpret the first
   * argument as the component to which the 
   * command has to be relayed, using this string
   * as a key.
   *
   * @return The compact name that identifies this
   * CLI
   */
  String getName();

  /**
   * Callback that gets called when a command line 
   * is parsed and associated to the implementing 
   * component.
   *
   * @param args The command line arguments.
   */
   void call(String[] args);
}
\end{lstlisting}

\subsubsection{SAMP Handlers} A possible extension that plug-ins can offer to
the users is SAMP handlers. When Iris receives a SAMP message that matches the
Handler's \verb|mtype|, the message is directly dispatched to the Handler itself
by the Iris framework. As a matter of fact, Iris just offers a convenient
shortcut to the excellent \verb|jsamp| implementation of SAMP, making it
available to the users with just the bare minimum amount of work required. The
setup of the SAMP infrastructure through \verb|jsamp| is all done by Iris,
including a keep-alive mechanism that brings a SAMP hub up when an existing one is
shut down.

A hook is provided for Components willing to send their own SAMP messages to the
SAMP Hub, again as a convenient shortcut to \verb|jsamp|.

\subsubsection{Custom Events} The Iris Events Framework is itself extensible:
this way plug-in developers can, if needed, create their own nested
architecture for their plug-in's Components.

\subsubsection{SED attachments} Components can attach arbitrary objects to the
SEDs managed by the SEDManager. This way users can rely on the Iris framework to
manage the additional information they might want to store about the individual
SEDs. When SEDs are deleted, the manager takes care of releasing any references
to the attachments, reducing the risk of memory leaks.

\subsection{Plug-in examples} \subsubsection{ASDC --- stable} The Italian Space
Agency Science Data Center (ASDC) hosts a database with tens of catalogs in a
very wide range of wavelengths, also providing time domain information.

A plug-in for providing Iris with a rich graphical user interface to query their
database was developed by the ASDC in a collaboration between the ASDC and the
Iris teams. The plug-in became part of the main Iris distribution in v2.0 and
was a valuable test bench to review, validate, and improve the Iris
Software Development Kit.

While the ASDC data query tool is now part of the Iris distribution, this tool
provides a very good example of how a plug-in can be integrated seamlessly in
the Iris framework to add specific value to the overall application.
Integration can be so seamless,
actually, that including the plug-in into the main Iris distribution is almost
exclusively a matter of configuration rather than of coding.

The ASDC data query tool extends the capabilities of the SED Builder by
providing a rich graphical user interface that allows users to check what
archives to query, and since the ASDC query is a positional cone search, the
client provides different adjustable search radii for each catalog that default
to reasonable values consistent with the resolving power of the individual
instruments.

Moreover, the tool allows users to query for specific observation time ranges, thus
allowing basic time domain analysis of the SEDs.

This component proves several points about the Iris framework and SDK, as listed below.
\begin{description} \begin{sloppypar}
\item[Custom-to-Standard adapters] The ASDC web service
backing up the implementation of the query tool does not comply with any VO data
access protocols (at least not yet), as this service was designed as a private interface
to their database to be consumed by a dedicated client like the one implemented in
Iris. The data files coming from the service, on the other hand, are compliant
with the IVOA specifications, so they can be directly read by SEDLib and passed
to the SEDManager.
  \item[Interoperability] Although not
designed as part of Iris, the ASDC plug-in integrates seamlessly with the Iris built-in
components. When the ASDC query tool downloads data from the service, the data
are listed in the SED Builder and visualized in the SED Viewer, even though the
ASDC tool
does not interact directly with any of them. They all interact only with the SED
Manager and they get notified of changes by the events that are fired when
Models are changed.\end{sloppypar}  \item[The Iris SDK] As it will be explored in some detail
in Section \ref{sec:writeplugin}, a plug-in developer can pretty much focus on
the implementation of the components' business logic, without worrying too much
about the boiler plate code required to configure such components. By using the
abstract classes that the Iris framework provides, one can leverage the existing
components with just a few lines of code and then start adding value to the
entire application.  \end{description}

\subsubsection{Vizier --- experimental} \label{sec:asdc} Experimental plug-ins
are shipped with Iris but they can only be activated by turning on switches on
the Iris command line. For instance, if one starts Iris with the command
\verb|./Iris --vizier| an experimental plug-in\footnote{While this client should work
fine most of the time,
users should not expect proper error handling, neither there is a way to
change the search radius.} for the CDS Vizier photometric
service gets loaded in the usual Iris desktop.

\subsubsection{R --- experimental} A highly experimental proof-of-concept
plug-in was developed to explore the possibility of interfacing Iris with rich
analysis environments like R. The plug-in shows how one can \emph{beam} data
from Iris to R and trigger some analysis on the dataset in R.\footnote{In order
to make this plug-in work one needs to install R and the Java-R interface package, and then
set up some environment variables and start Iris in a non-standard way.
If interested in experimenting with this plug-in,
please contact the corresponding author.}

\subsection{Other Extensibility Points}

\subsubsection{Custom File Readers} Iris supports a fair number of file formats
natively: VOTable, FITS, CSV, TSV, ASCII, and IPAC tables. However, new file
filters can be created and loaded at run-time. One can also create filters for
the natively supported files. In this case, the custom filter would parse the
file and map the metadata to the IVOA Data Model fields.

\subsubsection{Persistence} Components can also get a handle to the
configuration directory (usually a hidden folder in the user's home directory)
if they need to persist information like user's preferences, local databases, or
work sessions.

\subsection{How to write an Iris plug-in} \label{sec:writeplugin} Iris uses Maven
Archetypes to streamline the process of building and distributing Iris Java plug-ins.

You might also write plug-ins without using Maven, but you would need to take
care of many steps that the Maven-generated project automatically takes care of,
like the inclusion of your dependencies in your plug-in's jar file.

In order to have a test plug-in up and running you need to create a new project
from the Maven archetype:

\begin{lstlisting}[style=code]
$ export repo=http://vaotest2.tuc.noao.edu:8080/artifactory/

$ mvn archetype:generate\
  -DarchetypeRepository=$repo \
  -DarchetypeArtifactId=iris-plugin-archetype \
  -DarchetypeGroupId=cfa.vo \
  -DarchetypeVersion=1.1
\end{lstlisting}

The above command will ask you some questions about the metadata for your
plug-in project, like the group id, the project id (called artifact-id in
Maven), and the version. At the end of the process you should have a directory
named after your project-id. This directory contains all the files needed to
build and package a test plug-in.

You can type \verb|mvn package| from the newly created directory and Maven will
package the test plug-in for you in the \verb|target| directory as a jar file.

You can use the Iris Plug-in Manager component to install this jar file into
Iris. As soon as the plug-in is installed, a new button should appear on the Iris desktop. If
you click on the button, a rather impressive dialog box with the universal salutation
``Hello World!'' should appear on your screen.

You can inspect the source code of this project and notice that most of the code is
made of metadata strings and basic class definitions and instantiations. By
inheriting from the abstract classes that are provided with the Iris SDK, the
actual code that one needs to implement starts from the implementation of the
\verb|onClick| callback of the \verb|AbstractPluginMenuItem| class. From that call on,
a plug-in developer can focus on the implementation of their components and
start using the hooks provided by the Iris Framework in order to interoperate
with the other Iris components, and possibly with other VO applications.

One can start from this dummy project, inspect the source code, make changes to
the package and class names and to the metadata strings, and then start
implementing their component's business logic and user interface.

The Iris website contains further documentation on how to write plug-ins, and
you can contact the authors of this paper for further information.

\section{Future Plans} We are working on improving Iris in several ways. With
the VAO shutting down in 2014, the development of Iris has been taken over by
the Chandra X-Ray Center group at the Smithsonian Astrophysical Observatory.

While the current Software Development Kit is focused on letting plug-ins
contribute SEDs and SED segments to the user's workspace, we want to improve the
ways in which plug-ins can interact with the visualization and fitting code,
decoupling Specview and Sherpa.

We are also exploring solutions to overcome one of the limitations in the
current code, namely the handling of high resolution spectra, that is mostly due
to a visualization issue.

Several improvements will derive from the inclusion in Iris of the latest Sherpa
version, and in particular of the new code for interpolating templates in
template fitting. This will allow users to combine templates with other
templates and functions and compute photometric redshifts through template
fitting, for instance.

Also, we want to provide finer grained control over the visualization and
manipulation of individual components in the model expressions.

From the user interface point of view, we are planning to provide Python
bindings to enhance the integration of Iris in customized, complex scientific
work-flows.

\section{Conclusions} \label{sec:conclusions}

Iris is a Virtual Observatory application designed with the goal of streamlining
the construction of broadband spectral energy distributions while providing
flexible and robust tools for their analysis, with a stress on interoperability
and extensibility.

To summarize, Iris provides: built-in capabilities for building, viewing, and
analyzing broad-band spec\-tro-pho\-to\-met\-ric SEDs; a Python framework for fitting
user-pro\-vid\-ed models and templates; interoperability with Virtual Observatory
tools through the Simple Messaging Application Protocol (SAMP).

The Iris layered architecture takes advantage of the Virtual Observatory
standards and protocols without exposing their complexity to the end users,
who still benefit from the added interoperability. At the same time, developers
can use a middle layer of abstraction that exposes the domain objects, i.e.,
photometric SEDs, and the user's workspace, in a clean and consistent
way through a Java software development kit.

\begin{sloppypar}
This way Iris combines several existing software components with new dedicated
software, and provides hooks for astronomers and software developers that want to
leverage the general interoperable framework while plugging in their own code.
\end{sloppypar}

Iris is available as an Open Source project, and can be downloaded as a binary
or source distribution for Linux and OS X.

\section*{Acknowledgments}
\begin{sloppypar}
The Authors would like to acknowledge Giuseppina Fabbiano, Ian Evans,
Jonathan McDowell, and Aneta Siemiginowska
for their support and feedback in all the phases of the work.
Dan Nguyen and Joseph Miller (SAO) supported the
development team in the very early stages of the work.
\end{sloppypar}

We also thank the Italian Space Agency Science Data Center for the contribution
of the ASDC Data plug-in, in particular Paolo Giommi, Roberto Primavera,
Milvia Capalbi, and Bruce Gendre.

Support for the development of Iris was provided by
the Virtual Astronomical Observatory contract AST0834235. Support for Sherpa is
provided by the National Aeronautics and Space Administration through the
Chandra X-ray Center, which is operated by the Smithsonian Astrophysical
Observatory for and on behalf of the National Aeronautics and Space
Administration contract NAS8-03060.  Support for Specview is provided by the
Space Telescope Science Institute, operated by the Association of Universities
for Research in Astronomy, Inc., under National Aeronautics and Space
Administration contract NAS5-26555. This research has made use of the NASA/IPAC
Extragalactic Database (NED) which is operated by the Jet Propulsion Laboratory,
California Institute of Technology, under contract with the National Aeronautics
and Space Administration.


\begin{thebibliography}{47}
\expandafter\ifx\csname natexlab\endcsname\relax\def\natexlab#1{#1}\fi
\providecommand{\url}[1]{\texttt{#1}}
\providecommand{\href}[2]{#2}
\providecommand{\path}[1]{#1}
\providecommand{\DOIprefix}{doi:}
\providecommand{\ArXivprefix}{arXiv:}
\providecommand{\URLprefix}{URL: }
\providecommand{\Pubmedprefix}{pmid:}
\providecommand{\doi}[1]{\href{http://dx.doi.org/#1}{\path{#1}}}
\providecommand{\Pubmed}[1]{\href{pmid:#1}{\path{#1}}}
\providecommand{\bibinfo}[2]{#2}
\ifx\xfnm\relax \def\xfnm[#1]{\unskip,\space#1}\fi
\bibitem[{{Acquaviva} et~al.(2011){Acquaviva}, {Gawiser} and
  {Guaita}}]{2011ApJ...737...47A}
\bibinfo{author}{{Acquaviva}, V.}, \bibinfo{author}{{Gawiser}, E.},
  \bibinfo{author}{{Guaita}, L.}, \bibinfo{year}{2011}.
\newblock \bibinfo{title}{{Spectral Energy Distribution Fitting with Markov
  Chain Monte Carlo: Methodology and Application to z = 3.1
  Ly{$\alpha$}-emitting Galaxies}}.
\newblock \bibinfo{journal}{\apj} \bibinfo{volume}{737}, \bibinfo{pages}{47}.
\newblock \DOIprefix\doi{10.1088/0004-637X/737/2/47}.
\bibitem[{{Arnouts} et~al.(1999){Arnouts}, {Cristiani}, {Moscardini},
  {Matarrese}, {Lucchin}, {Fontana} and {Giallongo}}]{1999MNRAS.310..540A}
\bibinfo{author}{{Arnouts}, S.}, \bibinfo{author}{{Cristiani}, S.},
  \bibinfo{author}{{Moscardini}, L.}, \bibinfo{author}{{Matarrese}, S.},
  \bibinfo{author}{{Lucchin}, F.}, \bibinfo{author}{{Fontana}, A.},
  \bibinfo{author}{{Giallongo}, E.}, \bibinfo{year}{1999}.
\newblock \bibinfo{title}{{Measuring and modelling the redshift evolution of
  clustering: the Hubble Deep Field North}}.
\newblock \bibinfo{journal}{\mnras} \bibinfo{volume}{310},
  \bibinfo{pages}{540--556}.
\newblock \DOIprefix\doi{10.1046/j.1365-8711.1999.02978.x}.
\bibitem[{{Bayo} et~al.(2008){Bayo}, {Rodrigo}, {Barrado Y Navascu{\'e}s},
  {Solano}, {Guti{\'e}rrez}, {Morales-Calder{\'o}n} and
  {Allard}}]{2008A&A...492..277B}
\bibinfo{author}{{Bayo}, A.}, \bibinfo{author}{{Rodrigo}, C.},
  \bibinfo{author}{{Barrado Y Navascu{\'e}s}, D.}, \bibinfo{author}{{Solano},
  E.}, \bibinfo{author}{{Guti{\'e}rrez}, R.},
  \bibinfo{author}{{Morales-Calder{\'o}n}, M.}, \bibinfo{author}{{Allard}, F.},
  \bibinfo{year}{2008}.
\newblock \bibinfo{title}{{VOSA: virtual observatory SED analyzer. An
  application to the Collinder 69 open cluster}}.
\newblock \bibinfo{journal}{\aap} \bibinfo{volume}{492},
  \bibinfo{pages}{277--287}.
\newblock \DOIprefix\doi{10.1051/0004-6361:200810395}.
\bibitem[{{Ben{\'{\i}}tez}(2000)}]{2000ApJ...536..571B}
\bibinfo{author}{{Ben{\'{\i}}tez}, N.}, \bibinfo{year}{2000}.
\newblock \bibinfo{title}{{Bayesian Photometric Redshift Estimation}}.
\newblock \bibinfo{journal}{\apj} \bibinfo{volume}{536},
  \bibinfo{pages}{571--583}.
\newblock \DOIprefix\doi{10.1086/308947}.
\bibitem[{{Berriman} et~al.(2012){Berriman}, {Hanisch}, {Lazio}, {Szalay} and
  {Fabbiano}}]{2012SPIE.8449E..0HB}
\bibinfo{author}{{Berriman}, G.B.}, \bibinfo{author}{{Hanisch}, R.J.},
  \bibinfo{author}{{Lazio}, T.J.W.}, \bibinfo{author}{{Szalay}, A.},
  \bibinfo{author}{{Fabbiano}, G.}, \bibinfo{year}{2012}.
\newblock \bibinfo{title}{{The organization and management of the Virtual
  Astronomical Observatory}}, in: \bibinfo{booktitle}{Modeling, Systems
  Engineering, and Project Management for Astronomy V}.
\newblock \DOIprefix\doi{10.1117/12.926605}.
\bibitem[{{B{\l}a{\.z}ejowski} et~al.(2004){B{\l}a{\.z}ejowski},
  {Siemiginowska}, {Sikora}, {Moderski} and {Bechtold}}]{2004ApJ...600L..27B}
\bibinfo{author}{{B{\l}a{\.z}ejowski}, M.}, \bibinfo{author}{{Siemiginowska},
  A.}, \bibinfo{author}{{Sikora}, M.}, \bibinfo{author}{{Moderski}, R.},
  \bibinfo{author}{{Bechtold}, J.}, \bibinfo{year}{2004}.
\newblock \bibinfo{title}{{X-Ray Emission from the Quasar PKS 1127-145:
  Comptonized Infrared Photons on Parsec Scales}}.
\newblock \bibinfo{journal}{\apjl} \bibinfo{volume}{600},
  \bibinfo{pages}{L27--L30}.
\newblock \DOIprefix\doi{10.1086/381497}.
\bibitem[{{Bolzonella} et~al.(2000){Bolzonella}, {Miralles} and
  {Pell{\'o}}}]{2000A&A...363..476B}
\bibinfo{author}{{Bolzonella}, M.}, \bibinfo{author}{{Miralles}, J.M.},
  \bibinfo{author}{{Pell{\'o}}, R.}, \bibinfo{year}{2000}.
\newblock \bibinfo{title}{{Photometric redshifts based on standard SED fitting
  procedures}}.
\newblock \bibinfo{journal}{\aap} \bibinfo{volume}{363},
  \bibinfo{pages}{476--492}.
\newblock \href{http://arxiv.org/abs/astro-ph/0003380}{\tt
  arXiv:astro-ph/0003380}.
\bibitem[{{Budav{\'a}ri} et~al.(2009){Budav{\'a}ri}, {Wild}, {Szalay}, {Dobos}
  and {Yip}}]{2009MNRAS.394.1496B}
\bibinfo{author}{{Budav{\'a}ri}, T.}, \bibinfo{author}{{Wild}, V.},
  \bibinfo{author}{{Szalay}, A.S.}, \bibinfo{author}{{Dobos}, L.},
  \bibinfo{author}{{Yip}, C.W.}, \bibinfo{year}{2009}.
\newblock \bibinfo{title}{{Reliable eigenspectra for new generation surveys}}.
\newblock \bibinfo{journal}{\mnras} \bibinfo{volume}{394},
  \bibinfo{pages}{1496--1502}.
\newblock \DOIprefix\doi{10.1111/j.1365-2966.2009.14415.x}.
\bibitem[{{Busko}(2000)}]{2000ASPC..216...79B}
\bibinfo{author}{{Busko}, I.}, \bibinfo{year}{2000}.
\newblock \bibinfo{title}{{SPECVIEW: An Interactive Java Tool for Visualization
  and Analysis of Spectral Data}}, in: \bibinfo{booktitle}{Astronomical Data
  Analysis Software and Systems IX}, p.~\bibinfo{pages}{79}.
\bibitem[{{Busko}(2002)}]{2002ASPC..281..120B}
\bibinfo{author}{{Busko}, I.}, \bibinfo{year}{2002}.
\newblock \bibinfo{title}{{Specview: a Java Tool for Spectral Visualization and
  Model Fitting}}, in: \bibinfo{booktitle}{Astronomical Data Analysis Software
  and Systems XI}, p. \bibinfo{pages}{120}.
\bibitem[{{Cash}(1979)}]{1979ApJ...228..939C}
\bibinfo{author}{{Cash}, W.}, \bibinfo{year}{1979}.
\newblock \bibinfo{title}{{Parameter estimation in astronomy through
  application of the likelihood ratio}}.
\newblock \bibinfo{journal}{\apj} \bibinfo{volume}{228},
  \bibinfo{pages}{939--947}.
\newblock \DOIprefix\doi{10.1086/156922}.
\bibitem[{{Chen} et~al.(2005){Chen}, {Jura}, {Gordon} and
  {Blaylock}}]{2005ApJ...623..493C}
\bibinfo{author}{{Chen}, C.H.}, \bibinfo{author}{{Jura}, M.},
  \bibinfo{author}{{Gordon}, K.D.}, \bibinfo{author}{{Blaylock}, M.},
  \bibinfo{year}{2005}.
\newblock \bibinfo{title}{{A Spitzer Study of Dusty Disks in the
  Scorpius-Centaurus OB Association}}.
\newblock \bibinfo{journal}{\apj} \bibinfo{volume}{623},
  \bibinfo{pages}{493--501}.
\newblock \DOIprefix\doi{10.1086/428607}.
\bibitem[{{Chiang} and {Goldreich}(1997)}]{1997ApJ...490..368C}
\bibinfo{author}{{Chiang}, E.I.}, \bibinfo{author}{{Goldreich}, P.},
  \bibinfo{year}{1997}.
\newblock \bibinfo{title}{{Spectral Energy Distributions of T Tauri Stars with
  Passive Circumstellar Disks}}.
\newblock \bibinfo{journal}{\apj} \bibinfo{volume}{490},
  \bibinfo{pages}{368--376}.
\newblock \href{http://arxiv.org/abs/astro-ph/9706042}{\tt
  arXiv:astro-ph/9706042}.
\bibitem[{{Cid Fernandes} et~al.(2004){Cid Fernandes}, {Gu}, {Melnick},
  {Terlevich}, {Terlevich}, {Kunth}, {Rodrigues Lacerda} and
  {Joguet}}]{2004MNRAS.355..273C}
\bibinfo{author}{{Cid Fernandes}, R.}, \bibinfo{author}{{Gu}, Q.},
  \bibinfo{author}{{Melnick}, J.}, \bibinfo{author}{{Terlevich}, E.},
  \bibinfo{author}{{Terlevich}, R.}, \bibinfo{author}{{Kunth}, D.},
  \bibinfo{author}{{Rodrigues Lacerda}, R.}, \bibinfo{author}{{Joguet}, B.},
  \bibinfo{year}{2004}.
\newblock \bibinfo{title}{{The star formation history of Seyfert 2 nuclei}}.
\newblock \bibinfo{journal}{\mnras} \bibinfo{volume}{355},
  \bibinfo{pages}{273--296}.
\newblock \DOIprefix\doi{10.1111/j.1365-2966.2004.08321.x}.
\bibitem[{{Conroy}(2013)}]{2013ARA&A..51..393C}
\bibinfo{author}{{Conroy}, C.}, \bibinfo{year}{2013}.
\newblock \bibinfo{title}{{Modeling the Panchromatic Spectral Energy
  Distributions of Galaxies}}.
\newblock \bibinfo{journal}{\araa} \bibinfo{volume}{51},
  \bibinfo{pages}{393--455}.
\newblock \DOIprefix\doi{10.1146/annurev-astro-082812-141017}.
\bibitem[{{Czerny} and {Elvis}(1987)}]{1987ApJ...321..305C}
\bibinfo{author}{{Czerny}, B.}, \bibinfo{author}{{Elvis}, M.},
  \bibinfo{year}{1987}.
\newblock \bibinfo{title}{{Constraints on quasar accretion disks from the
  optical/ultraviolet/soft X-ray big bump}}.
\newblock \bibinfo{journal}{\apj} \bibinfo{volume}{321},
  \bibinfo{pages}{305--320}.
\newblock \DOIprefix\doi{10.1086/165630}.
\bibitem[{{Dermer} and {Schlickeiser}(2002)}]{2002ApJ...575..667D}
\bibinfo{author}{{Dermer}, C.D.}, \bibinfo{author}{{Schlickeiser}, R.},
  \bibinfo{year}{2002}.
\newblock \bibinfo{title}{{Transformation Properties of External Radiation
  Fields, Energy-Loss Rates and Scattered Spectra, and a Model for Blazar
  Variability}}.
\newblock \bibinfo{journal}{\apj} \bibinfo{volume}{575},
  \bibinfo{pages}{667--686}.
\newblock \DOIprefix\doi{10.1086/341431}.
\bibitem[{{Doe} et~al.(2012)}]{2012ASPC..461..893D}
\bibinfo{author}{{Doe}, S.}, et~al., \bibinfo{year}{2012}.
\newblock \bibinfo{title}{{Iris: The VAO SED Application}}, in:
  \bibinfo{booktitle}{Astronomical Data Analysis Software and Systems XXI}, p.
  \bibinfo{pages}{893}.
\newblock \href{http://arxiv.org/abs/1205.2419}{\tt arXiv:1205.2419}.
\bibitem[{{Evans} et~al.(2012)}]{2012SPIE.8449E..0IE}
\bibinfo{author}{{Evans}, J.D.}, et~al., \bibinfo{year}{2012}.
\newblock \bibinfo{title}{{Managing distributed software development in the
  Virtual Astronomical Observatory}}, in: \bibinfo{booktitle}{Modeling, Systems
  Engineering, and Project Management for Astronomy V}.
\newblock \DOIprefix\doi{10.1117/12.927371}.
\bibitem[{{Freeman} et~al.(2001){Freeman}, {Doe} and
  {Siemiginowska}}]{2001SPIE.4477...76F}
\bibinfo{author}{{Freeman}, P.}, \bibinfo{author}{{Doe}, S.},
  \bibinfo{author}{{Siemiginowska}, A.}, \bibinfo{year}{2001}.
\newblock \bibinfo{title}{{Sherpa: a mission-independent data analysis
  application}}, in: \bibinfo{booktitle}{Astronomical Data Analysis}, pp.
  \bibinfo{pages}{76--87}.
\newblock \DOIprefix\doi{10.1117/12.447161}.
\bibitem[{{Fruscione} et~al.(2006)}]{2006SPIE.6270E..60F}
\bibinfo{author}{{Fruscione}, A.}, et~al., \bibinfo{year}{2006}.
\newblock \bibinfo{title}{{CIAO: Chandra's data analysis system}}, in:
  \bibinfo{booktitle}{Observatory Operations: Strategies, Processes, and
  Systems}.
\newblock \DOIprefix\doi{10.1117/12.671760}.
\bibitem[{{Ilbert} et~al.(2006)}]{2006A&A...457..841I}
\bibinfo{author}{{Ilbert}, O.}, et~al., \bibinfo{year}{2006}.
\newblock \bibinfo{title}{{Accurate photometric redshifts for the CFHT legacy
  survey calibrated using the VIMOS VLT deep survey}}.
\newblock \bibinfo{journal}{\aap} \bibinfo{volume}{457},
  \bibinfo{pages}{841--856}.
\newblock \DOIprefix\doi{10.1051/0004-6361:20065138}.
\bibitem[{{Johnson} and {Foote}(1988)}]{ioc}
\bibinfo{author}{{Johnson}, R.E.}, \bibinfo{author}{{Foote}, B.},
  \bibinfo{year}{1988}.
\newblock \bibinfo{title}{{Designing Reusable Classes}}.
\newblock \bibinfo{journal}{{Journal of Object-Oriented Programming}}
  \bibinfo{volume}{1}, \bibinfo{pages}{22--35}.
\newblock \bibinfo{note}{\url{http://www.laputan.org/drc.html}}.
\bibitem[{{Lagarias} et~al.(1998){Lagarias}, {Reeds}, {Wright} and
  {Wright}}]{Lagarias:1998:CPN:588893.589108}
\bibinfo{author}{{Lagarias}, J.C.}, \bibinfo{author}{{Reeds}, J.A.},
  \bibinfo{author}{{Wright}, M.H.}, \bibinfo{author}{{Wright}, P.E.},
  \bibinfo{year}{1998}.
\newblock \bibinfo{title}{{Convergence Properties of the Nelder--Mead Simplex
  Method in Low Dimensions}}.
\newblock \bibinfo{journal}{SIAM J. on Optimization} \bibinfo{volume}{9},
  \bibinfo{pages}{112--147}.
\newblock \DOIprefix\doi{10.1137/S1052623496303470}.
\bibitem[{{Lagrange} et~al.(2000){Lagrange}, {Backman} and
  {Artymowicz}}]{2000prpl.conf..639L}
\bibinfo{author}{{Lagrange}, A.M.}, \bibinfo{author}{{Backman}, D.E.},
  \bibinfo{author}{{Artymowicz}, P.}, \bibinfo{year}{2000}.
\newblock \bibinfo{title}{{Planetary Material around Main-Sequence Stars}}.
\newblock \bibinfo{journal}{Protostars and Planets IV} , \bibinfo{pages}{639}.
\bibitem[{{Laurino} et~al.(2013){Laurino}, {Busko}, {Cresitello-Dittmar},
  {D'Abrusco}, {Doe}, {Evans}, {Pevunova} and {Norris}}]{2013AAS...22124038L}
\bibinfo{author}{{Laurino}, O.}, \bibinfo{author}{{Busko}, I.},
  \bibinfo{author}{{Cresitello-Dittmar}, M.}, \bibinfo{author}{{D'Abrusco},
  R.}, \bibinfo{author}{{Doe}, S.}, \bibinfo{author}{{Evans}, J.},
  \bibinfo{author}{{Pevunova}, O.}, \bibinfo{author}{{Norris}, P.},
  \bibinfo{year}{2013}.
\newblock \bibinfo{title}{{Constructing and Analyzing Spectral Energy
  Distributions with the Virtual Observatory}}, in:
  \bibinfo{booktitle}{American Astronomical Society Meeting Abstracts}, p.
  \bibinfo{pages}{240.38}.
\bibitem[{{Massaro} et~al.(2006){Massaro}, {Tramacere}, {Perri}, {Giommi} and
  {Tosti}}]{2006A&A...448..861M}
\bibinfo{author}{{Massaro}, E.}, \bibinfo{author}{{Tramacere}, A.},
  \bibinfo{author}{{Perri}, M.}, \bibinfo{author}{{Giommi}, P.},
  \bibinfo{author}{{Tosti}, G.}, \bibinfo{year}{2006}.
\newblock \bibinfo{title}{{Log-parabolic spectra and particle acceleration in
  blazars. III. SSC emission in the TeV band from Mkn501}}.
\newblock \bibinfo{journal}{\aap} \bibinfo{volume}{448},
  \bibinfo{pages}{861--871}.
\newblock \DOIprefix\doi{10.1051/0004-6361:20053644}.
\bibitem[{{Mazzarella} and {NED Team}(2007)}]{2007ASPC..376..153M}
\bibinfo{author}{{Mazzarella}, J.M.}, \bibinfo{author}{{NED Team}},
  \bibinfo{year}{2007}.
\newblock \bibinfo{title}{{NED for a New Era}}, in:
  \bibinfo{booktitle}{Astronomical Data Analysis Software and Systems XVI}, p.
  \bibinfo{pages}{153}.
\bibitem[{{McDowell} et~al.(2012)}]{2012arXiv1204.3055M}
\bibinfo{author}{{McDowell}, J.}, et~al., \bibinfo{year}{2012}.
\newblock \bibinfo{title}{{IVOA Recommendation: Spectrum Data Model 1.1}}.
\newblock \bibinfo{journal}{ArXiv e-prints}
  \href{http://arxiv.org/abs/1204.3055}{\tt arXiv:1204.3055}.
\bibitem[{{Moré}(1978)}]{jjmore1978}
\bibinfo{author}{{Moré}, J.J.}, \bibinfo{year}{1978}.
\newblock \bibinfo{title}{The levenberg-marquardt algorithm: Implementation and
  theory}, in: \bibinfo{editor}{Watson, G.} (Ed.),
  \bibinfo{booktitle}{Numerical Analysis}. \bibinfo{publisher}{Springer Berlin
  Heidelberg}. volume \bibinfo{volume}{630} of \textit{\bibinfo{series}{Lecture
  Notes in Mathematics}}, pp. \bibinfo{pages}{105--116}.
\newblock \DOIprefix\doi{10.1007/BFb0067700}.
\bibitem[{{Ochsenbein} et~al.(2011)}]{2011arXiv1110.0524O}
\bibinfo{author}{{Ochsenbein}, F.}, et~al., \bibinfo{year}{2011}.
\newblock \bibinfo{title}{{IVOA Recommendation: VOTable Format Definition
  Version 1.2}}.
\newblock \bibinfo{journal}{ArXiv e-prints}
  \href{http://arxiv.org/abs/1110.0524}{\tt arXiv:1110.0524}.
\bibitem[{{Quinn} et~al.(2004)}]{2004SPIE.5493..137Q}
\bibinfo{author}{{Quinn}, P.J.}, et~al., \bibinfo{year}{2004}.
\newblock \bibinfo{title}{{The International Virtual Observatory Alliance:
  recent technical developments and the road ahead}}, in:
  \bibinfo{booktitle}{Optimizing Scientific Return for Astronomy through
  Information Technologies}, pp. \bibinfo{pages}{137--145}.
\newblock \DOIprefix\doi{10.1117/12.551247}.
\bibitem[{{Refsdal} et~al.(2009)}]{2009pysc.conf...51R}
\bibinfo{author}{{Refsdal}, B.L.}, et~al., \bibinfo{year}{2009}.
\newblock \bibinfo{title}{{Sherpa: 1D/2D modeling and fitting in Python}}, in:
  \bibinfo{booktitle}{Proceedings of the 8th Python in Science Conference,
  Pasadena, CA, 2009, edited by G. Varoquaux, S. van der Walt and J. Millman},
  p.~\bibinfo{pages}{51}.
\bibitem[{{Robitaille} et~al.(2007){Robitaille}, {Whitney}, {Indebetouw} and
  {Wood}}]{2007ApJS..169..328R}
\bibinfo{author}{{Robitaille}, T.P.}, \bibinfo{author}{{Whitney}, B.A.},
  \bibinfo{author}{{Indebetouw}, R.}, \bibinfo{author}{{Wood}, K.},
  \bibinfo{year}{2007}.
\newblock \bibinfo{title}{{Interpreting Spectral Energy Distributions from
  Young Stellar Objects. II. Fitting Observed SEDs Using a Large Grid of
  Precomputed Models}}.
\newblock \bibinfo{journal}{\apjs} \bibinfo{volume}{169},
  \bibinfo{pages}{328--352}.
\newblock \DOIprefix\doi{10.1086/512039}.
\bibitem[{{Robitaille} et~al.(2006){Robitaille}, {Whitney}, {Indebetouw},
  {Wood} and {Denzmore}}]{2006ApJS..167..256R}
\bibinfo{author}{{Robitaille}, T.P.}, \bibinfo{author}{{Whitney}, B.A.},
  \bibinfo{author}{{Indebetouw}, R.}, \bibinfo{author}{{Wood}, K.},
  \bibinfo{author}{{Denzmore}, P.}, \bibinfo{year}{2006}.
\newblock \bibinfo{title}{{Interpreting Spectral Energy Distributions from
  Young Stellar Objects. I. A Grid of 200,000 YSO Model SEDs}}.
\newblock \bibinfo{journal}{\apjs} \bibinfo{volume}{167},
  \bibinfo{pages}{256--285}.
\newblock \DOIprefix\doi{10.1086/508424}.
\bibitem[{{Sawicki} and {Yee}(1998)}]{1998AJ....115.1329S}
\bibinfo{author}{{Sawicki}, M.}, \bibinfo{author}{{Yee}, H.K.C.},
  \bibinfo{year}{1998}.
\newblock \bibinfo{title}{{Optical-Infrared Spectral Energy Distributions of Z
  > 2 Lyman Break Galaxies}}.
\newblock \bibinfo{journal}{\aj} \bibinfo{volume}{115},
  \bibinfo{pages}{1329--1339}.
\newblock \DOIprefix\doi{10.1086/300291}.
\bibitem[{{Shapley} et~al.(2001){Shapley}, {Fabbiano} and
  {Eskridge}}]{2001ApJS..137..139S}
\bibinfo{author}{{Shapley}, A.}, \bibinfo{author}{{Fabbiano}, G.},
  \bibinfo{author}{{Eskridge}, P.B.}, \bibinfo{year}{2001}.
\newblock \bibinfo{title}{{A Multivariate Statistical Analysis of Spiral Galaxy
  Luminosities. I. Data and Results}}.
\newblock \bibinfo{journal}{\apjs} \bibinfo{volume}{137},
  \bibinfo{pages}{139--199}.
\newblock \DOIprefix\doi{10.1086/322998}.
\bibitem[{{Smith} et~al.(2007)}]{2007ApJ...656..770S}
\bibinfo{author}{{Smith}, J.D.T.}, et~al., \bibinfo{year}{2007}.
\newblock \bibinfo{title}{{The Mid-Infrared Spectrum of Star-forming Galaxies:
  Global Properties of Polycyclic Aromatic Hydrocarbon Emission}}.
\newblock \bibinfo{journal}{\apj} \bibinfo{volume}{656},
  \bibinfo{pages}{770--791}.
\newblock \DOIprefix\doi{10.1086/510549}.
\bibitem[{{Solano}(2013)}]{2013arXiv1312.3249S}
\bibinfo{author}{{Solano}, E.}, \bibinfo{year}{2013}.
\newblock \bibinfo{title}{{Spectral stellar libraries and the Virtual
  Observatory}}.
\newblock \bibinfo{journal}{ArXiv e-prints}
  \href{http://arxiv.org/abs/1312.3249}{\tt arXiv:1312.3249}.
\bibitem[{Storn and Price(1997)}]{Storn:1997:DEN:596061.596146}
\bibinfo{author}{Storn, R.}, \bibinfo{author}{Price, K.}, \bibinfo{year}{1997}.
\newblock \bibinfo{title}{Differential evolution \&ndash; a simple and
  efficient heuristic for global optimization over continuous spaces}.
\newblock \bibinfo{journal}{J. of Global Optimization} \bibinfo{volume}{11},
  \bibinfo{pages}{341--359}.
\newblock \DOIprefix\doi{10.1023/A:1008202821328}.
\bibitem[{{Taylor} et~al.(2011){Taylor}, {Boch}, {Fitzpatrick}, {Allan},
  {Paioro}, {Taylor} and {Tody}}]{2011arXiv1110.0528T}
\bibinfo{author}{{Taylor}, M.}, \bibinfo{author}{{Boch}, T.},
  \bibinfo{author}{{Fitzpatrick}, M.}, \bibinfo{author}{{Allan}, A.},
  \bibinfo{author}{{Paioro}, L.}, \bibinfo{author}{{Taylor}, J.},
  \bibinfo{author}{{Tody}, D.}, \bibinfo{year}{2011}.
\newblock \bibinfo{title}{{IVOA Recommendation: SAMP - Simple Application
  Messaging Protocol Version 1.3}}.
\newblock \bibinfo{journal}{ArXiv e-prints}
  \href{http://arxiv.org/abs/1110.0528}{\tt arXiv:1110.0528}.
\bibitem[{{Taylor}(2005)}]{2005ASPC..347...29T}
\bibinfo{author}{{Taylor}, M.B.}, \bibinfo{year}{2005}.
\newblock \bibinfo{title}{{TOPCAT {\&} STIL: Starlink Table/VOTable Processing
  Software}}, in: \bibinfo{booktitle}{Astronomical Data Analysis Software and
  Systems XIV}, p.~\bibinfo{pages}{29}.
\bibitem[{{Tody} et~al.(2012)}]{2012arXiv1203.5725T}
\bibinfo{author}{{Tody}, D.}, et~al., \bibinfo{year}{2012}.
\newblock \bibinfo{title}{{IVOA Recommendation: Simple Spectral Access Protocol
  Version 1.1}}.
\newblock \bibinfo{journal}{ArXiv e-prints}
  \href{http://arxiv.org/abs/1203.5725}{\tt arXiv:1203.5725}.
\bibitem[{{Tramacere} et~al.(2009){Tramacere}, {Giommi}, {Perri}, {Verrecchia}
  and {Tosti}}]{2009A&A...501..879T}
\bibinfo{author}{{Tramacere}, A.}, \bibinfo{author}{{Giommi}, P.},
  \bibinfo{author}{{Perri}, M.}, \bibinfo{author}{{Verrecchia}, F.},
  \bibinfo{author}{{Tosti}, G.}, \bibinfo{year}{2009}.
\newblock \bibinfo{title}{{Swift observations of the very intense flaring
  activity of Mrk 421 during 2006. I. Phenomenological picture of electron
  acceleration and predictions for MeV/GeV emission}}.
\newblock \bibinfo{journal}{\aap} \bibinfo{volume}{501},
  \bibinfo{pages}{879--898}.
\newblock \DOIprefix\doi{10.1051/0004-6361/200810865}.
\bibitem[{{Vrtilek} et~al.(1990){Vrtilek}, {Raymond}, {Garcia}, {Verbunt},
  {Hasinger} and {Kurster}}]{1990A&A...235..162V}
\bibinfo{author}{{Vrtilek}, S.D.}, \bibinfo{author}{{Raymond}, J.C.},
  \bibinfo{author}{{Garcia}, M.R.}, \bibinfo{author}{{Verbunt}, F.},
  \bibinfo{author}{{Hasinger}, G.}, \bibinfo{author}{{Kurster}, M.},
  \bibinfo{year}{1990}.
\newblock \bibinfo{title}{{Observations of Cygnus X-2 with IUE - Ultraviolet
  results from a multiwavelength campaign}}.
\newblock \bibinfo{journal}{\aap} \bibinfo{volume}{235},
  \bibinfo{pages}{162--173}.
\bibitem[{{Walcher} et~al.(2011){Walcher}, {Groves}, {Budav{\'a}ri} and
  {Dale}}]{2011Ap&SS.331....1W}
\bibinfo{author}{{Walcher}, J.}, \bibinfo{author}{{Groves}, B.},
  \bibinfo{author}{{Budav{\'a}ri}, T.}, \bibinfo{author}{{Dale}, D.},
  \bibinfo{year}{2011}.
\newblock \bibinfo{title}{{Fitting the integrated spectral energy distributions
  of galaxies}}.
\newblock \bibinfo{journal}{\apss} \bibinfo{volume}{331},
  \bibinfo{pages}{1--52}.
\newblock \DOIprefix\doi{10.1007/s10509-010-0458-z}.
\bibitem[{Wright(1996)}]{wright1996}
\bibinfo{author}{Wright, M.H.}, \bibinfo{year}{1996}.
\newblock \bibinfo{title}{{Direct Search Methods: Once Scorned, Now
  Respectable}}, in: \bibinfo{editor}{Griffiths, D.F.},
  \bibinfo{editor}{Watson, G.A.} (Eds.), \bibinfo{booktitle}{Numerical Analysis
  1995 (Proceedings of the 1995 Dundee Biennial Conference in Numerical
  Analysis)}, \bibinfo{publisher}{{CRC} Press}. pp. \bibinfo{pages}{191--208}.

\end{thebibliography}
\end{document}